\documentclass{emulateapj}
\usepackage{times}

\newcommand{\cmsq}{\mbox{ cm$^{2}$}}
\newcommand{\Mpc}{\mbox{ Mpc}}
\newcommand{\Mpcden}{\mbox{ Mpc}^{-3}}
\newcommand{\kpc}{\mbox{ kpc}}

\newcommand{\Mpcinv}{\mbox{ Mpc$^{-1}$}}

\newcommand{\kel}{\mbox{ K}}
\newcommand{\mkel}{\mbox{ mK}}

\newcommand{\yr}{\mbox{ yr}}

\newcommand{\secinv}{\mbox{ s$^{-1}$}}

\newcommand{\Msun}{\mbox{ M$_\odot$}}

\newcommand{\Lsun}{\mbox{ L$_\odot$}}

\newcommand{\hunits}{\mbox{ km s$^{-1}$ Mpc$^{-1}$}}

\newcommand{\kms}{\mbox{ km s$^{-1}$}}

\newcommand{\Junits}{\mbox{ cm$^{-2}$ s$^{-1}$ Hz$^{-1}$ sr$^{-1}$}}

\newcommand{\bxhi}{\bar{x}_{\rm HI}}
\newcommand{\xhi}{x_{\rm HI}}

\newcommand{\bxion}{\bar{x}_i}
\newcommand{\xion}{x_i}

\newcommand{\hone}{HI }
\newcommand{\htwo}{HII }
\newcommand{\mmin}{m_{\rm min}}
\newcommand{\fcoll}{f_{\rm coll}}

\newcommand{\lya}{Ly$\alpha$ }

\newcommand{\deriv}{{\rm d}}
\newcommand{\bq}{\begin{equation}}
\newcommand{\eq}{\end{equation}}
\newcommand{\bqa}{\begin{eqnarray}}
\newcommand{\eqa}{\end{eqnarray}}
\def\VEV#1{\left\langle #1\right\rangle} 

\newcommand\lsim{\mathrel{\rlap{\lower4pt\hbox{\hskip1pt$\sim$}}
        \raise1pt\hbox{$<$}}}
\newcommand\gsim{\mathrel{\rlap{\lower4pt\hbox{\hskip1pt$\sim$}}
        \raise1pt\hbox{$>$}}}

\begin{document}

\title{Fossil Ionized Bubbles Around Dead Quasars During Reionization}

\author{Steven R.  Furlanetto,\altaffilmark{1} Zolt\'{a}n Haiman,\altaffilmark{2} \& S.~Peng Oh\altaffilmark{3}}

\altaffiltext{1} {Department of Physics and Astronomy, University of California, Los Angeles, CA 90095, USA; sfurlane@astro.ucla.edu}

\altaffiltext{2} {Department of Astronomy, Columbia University,  550 West 120th Street, New York, NY 10027, USA; zoltan@astro.columbia.edu}

\altaffiltext{3}{Department of Physics, University of California, Santa Barbara, CA 93106, USA; peng@physics.ucsb.edu}

\begin{abstract}
One of the most dramatic signatures of the reionization era may be the
enormous ionized bubbles around luminous quasars (with radii reaching
$\sim 40$ comoving Mpc), which may survive as ``fossil'' ionized
regions long after their source shuts off.  Here we study how the
inhomogeneous intergalactic medium (IGM) evolves inside such
fossils.  The average recombination rate declines rapidly with time:
the densest pockets recombine rapidly, leaving low--density regions
that recombine much more slowly.  Furthermore, the brief quasar episode
significantly increases the mean free path inside the fossil bubbles.
As a result, even a weak ionizing background generated by galaxies inside the
fossil can maintain it in a relatively highly and uniformly ionized
state. For example, galaxies that would ionize $20$--$30\%$ of hydrogen
in a random patch of the IGM can maintain $80$--$90\%$ ionization inside the fossil, for a duration much longer than the average recombination
time in the IGM.  Quasar fossils at $z\lsim 10$ can thus retain their
identity for nearly a Hubble time, and will appear ``gray,''
distinct from both the average IGM (which has a ``swiss-cheese"
ionization topology and a lower mean ionized fraction), and from
bubbles around active quasars (which are fully ionized).  More distant
fossils, at $z\gsim 10$ have a weaker galaxy-generated ionizing background and a higher gas density.  They can break up and attain a swiss-cheese topology similar to the rest of the IGM, but still with a
smaller contrast between the ionized bubbles and the partially neutral
regions separating them.  Analogous HeIII-fossils should exist around
the epoch of HeII/HeIII reionization at $z \sim 3$.  Rapid
recombinations inside the HeIII-fossils will be more common, because
many of them will have no HeII-ionizing background; but the time lag before another quasar appears is typically small, so $\la 50\%$ of the gas is able to recombine, even for fossils that form well before reionization is complete.  Our model of inhomogeneous recombination also applies to ``double reionization'' models and shows
that a non-monotonic reionization history is even more unlikely than previously thought.
\end{abstract}
  
\keywords{cosmology: theory -- intergalactic medium}

\section{Introduction} \label{intro}

The reionization of hydrogen and helium throughout the intergalactic
medium (IGM) are landmark events in the early history of structure
formation.  As such, they (and particularly hydrogen reionization)
have received a great deal of attention -- both observationally and
theoretically -- in the past several years.  At present, the
observational evidence leaves significant ambiguities (see
\citealt{fan06-review} for a recent review).  The electron scattering
optical depth of $\tau\sim 0.09$ measured through cosmic microwave
background polarization anisotropies implies that hydrogen
reionization began at $z \gsim 10$, albeit with a large uncertainty \citep{page06, dunkley08}.  
On the other hand, \lya absorption spectra of quasars
at $z \sim 6$ show some evidence for a rapid transition in the
globally-averaged neutral fraction, $\bxhi$ (e.g., \citealt{fan06}).
However the \lya absorption is so saturated in the \citet{gunn65}
trough (with optical depth $\tau_{\rm GP} \ga 10^5 \bxhi$) that
constraints derived directly from the lack of flux in that spectral
region are weak \citep{white03, fan06}.  Studying the properties of
the ionized zones and the flux transmitted at the corresponding
wavelengths near the \lya line has yielded much tighter lower limits
on the neutral fraction along a few sight lines, using \lya and Ly$\beta$
transmission statistics \citep{mesinger04,mesinger07-prox} and the inferred ionized
zone sizes \citep{wyithe04-prox, mesinger04}, although the interpretations of
these results is still subject to poor statistics and possible biases
(e.g, \citealt{lidz06, becker07,maselli07,bolton07,bolton07-prox}).  Helium
reionization is equally controversial: although HeII \lya forest
spectra suggest rapid evolution at $z \sim 2.9$ \citep{heap00,
smette02}, secondary signatures (such as the IGM thermal history and
radiation background) suggest change at slightly higher redshifts --
or in some cases no evolution at all (e.g., \citealt{schaye00,
theuns02-reion, fauch07, songaila98, kim02, aguirre04, songaila05, bernardi03}).

One of the most interesting aspect of the reionization process is its
inhomogeneity: different regions of the Universe can be reionized at
very different times, depending on their relative proximity to the
ionizing sources.  During reionization, the resulting patterns of
ionized and neutral gas can then be a rich source of information about
the IGM, the ionizing sources, and the ways they interact
\citep{furl04-bub, furl07-helium}.  In essence, large ionized zones
pinpoint either large clusters of sources or exceptionally bright
ones.

The most dramatic examples are luminous quasars, which can ionize
enormous regions, spanning several tens of (comoving) Mpc, in either
hydrogen or helium.  These sources are (probably) rather short-lived,
with expected lifetimes $\sim 10^7$--$10^8 \yr$ (see, e.g.,
\citealt{martini04} for a review). When the quasar shuts down, the
ionizing background within the region drops precipitously: the
galaxies that surround the quasar host may be relatively numerous
\citep{alvarez07, lidz07, geil07} but generically cannot compete with
such a strong ionizing source.  As such, the IGM will begin
recombining as soon as it shuts off.

However, the recombination time of IGM gas (at the mean density) is
comparable to the Hubble time at $z \sim 8$ and $z \sim 3$ for
hydrogen and helium, respectively.
Thus during either recombination era one expects the ionized region to
survive for a relatively long time after its source vanishes, becoming
a ``fossil'' bubble.  As some of the largest coherent features in the
Universe at these high redshifts, these bubbles (both during their
active and fossil phases) are obvious objects of interest.  In
particular, during hydrogen reionization they are favored targets of
first-generation 21 cm surveys, which can only make images on the
largest scales -- and so are limited to search for structures of these
sizes \citep{wyithe04-qso, wyithe05-qso}.  The characteristics of
these regions can be used to study the properties of the quasar
\citep{wyithe05-qso, zaroubi05, thomas08, kramer07,sethi08} as well as
its relation to the surrounding sources \citep{alvarez07, lidz07,
geil07}.  Although active bright quasars are quite rare at high
redshifts, the long lifetime of fossils will make large features
much easier to find -- increasing their abundance by approximately the
ratio of the recombination time to the source lifetime.  During helium
reionization, the distribution of these fossils will affect the
thermal histories of IGM gas elements and hence their \lya forest
properties \citep{gleser05, furl07-igmtemp}.

Despite their promise, there has been relatively little attention paid
to the physics of the fossils, and in particular to how they actually
recombine.  The most naive picture, in which the IGM is assumed to be
uniform at the mean density, is obviously too simple because, even at
these high redshifts, the filamentary structure of the cosmic web is
growing -- and it has long been recognized that the accompanying
clumping is crucial for understanding the reionization process.  In
that context, the most common approximation for this inhomogeneity is
the clumping factor $C= \VEV{n^2}/\VEV{n}^2$, which is the ratio of
the recombination rate (volume-averaged over all ionized regions) to
its naive expectation in a uniform medium.\footnote{In practice, it is
often approximated by averaging over the entire volume, excluding only
regions bound to dark matter halos (e.g., \citealt{iliev05-sim,
mellema06}), because weighting by the 
ionized fraction introduces extra complexity (but see \citealt{gnedin97,kohler05-sim}).}

This simple approach can be adequate so long as the spatial averaging
to derive the clumping factor excludes neutral regions, which are of
course not recombining, while simultaneously accounting for the
large--scale structure in the IGM \citep{miralda00, furl05-rec}.  In
that case, so long as the division of the IGM into neutral and ionized
regions is not changing rapidly, a constant clumping factor approach
is useful because the densest ionized regions (which account for most
of the recombinations) are relatively stable.

However, with fossil bubbles this assumption is manifestly false:
after such a precipitous drop in the ionizing background, the entire
region falls out of ionization equilibrium.  The dense gas will
initially be highly ionized by the quasar, so the clumping factor will
be large.  But these regions recombine quickly and become neutral,
making the total recombination rate (and effective clumping factor)
fall, eventually flattening out at a level appropriate to the smaller
ionizing background from the remaining galaxies.

Similar physics should describe any global ``recombination era'',
during which the global average ionized fraction is decreasing with
cosmic time.  This may occur in ``double reionization'' scenarios, in
which there is a sharp transition in the mode of star formation so
that the IGM is reionized twice, with substantial recombinations
in between \citep{cen03, cen03-letter, wyithe03, wyithe03-letter},
although in practice such histories are difficult to arrange
self-consistently \citep{haiman03, furl05-double, iliev07-selfreg}.
In this case, the entire IGM would be ``fossilized'', with the second
generation of stars unable to maintain the same level of ionization as
the previous one.  Such models have only been studied using a simple
clumping prescription, and they rely on a relatively large value for
$C$ to speed up recombinations during the intermediate phase
\citep{furl05-double}.  But the same arguments apply as for fossils: a
large clumping factor will only be appropriate until the dense gas
recombines.  Afterward, the effective clumping factor will
decrease, and the global recombination rate will slow down.

The goal of this paper is to describe the physics of this
recombination process in some detail and thus to evaluate better the
structure of fossil ionized regions.  We will aim to answer several
basic questions. Does the clumpiness of the IGM allow most of the gas
to recombine, or only the densest regions?  Can the residual ionizing
background from galaxies maintain high ionization, even if they could
not provide the initial ionization of the gas on their own?  How does
the ionization topology evolve inside the fossils? Does the
inhomogeneous IGM allow regions far from galaxies to shield themselves
and recombine faster than those near the galaxies -- so that the
fossil resembles the swiss--cheese topology of the rest of the IGM?
Or, does the ionization remain more uniform inside fossils than
outside?  And, finally, can quasar fossils be distinguished, either
individually or statistically, from large, galaxy-generated ionized
bubbles, and from active quasars, through the topology of their
ionized gas and their abundance?

This paper is organized as follows.  In \S~\ref{ionhist} and
\ref{mhr}, we describe our models for computing the recombination
histories of gas parcels and the density distribution of the IGM.
Then, in \S \ref{zeroion}--\ref{inhom}, we apply this model to fossils
produced during hydrogen reionization, studying, respectively, a fossil
with no ionizing sources, a fossil with a uniform ionizing background flux, 
and an alternative, ``global'' fossil model, incorporating the inhomogeneous
density field of the IGM, but explicitly conserving photon number.
In \S~\ref{helium}, we apply these models to fossils during helium
reionization.  Finally, in \S~\ref{disc}, we discuss some of the
observational prospects to detect these fossils, and we offer our
conclusions in \S~\ref{summ}.

In our numerical calculations, we assume a cosmology with
$\Omega_m=0.26$, $\Omega_\Lambda=0.74$, $\Omega_b=0.044$, $H_0=100 h
\hunits$ (with $h=0.74$), $n=0.95$, and $\sigma_8=0.8$, consistent
with the most recent measurements \citep{dunkley08,komatsu08}.  
Unless otherwise specified, we use comoving units for all distances.

\section{The Recombination History} \label{ionhist}

Our first task is to compute the recombination history of an individual gas parcel.  We will use a simplified version of \citet{hui97}; see \citet{furl07-igmtemp} for more details.  The full solution requires us to follow the abundances of three independent species (including hydrogen, helium, and electrons), the gas temperature (which affects the recombination and collisional ionization rates), and the density of the parcel.

Consider a gas element of fractional overdensity $\delta$ and temperature $T$ illuminated by an ionizing background of the form
\bq
J_\nu = J_{{\rm HI},-21} \left( {\nu \over \nu_{\rm HI}} \right)^{-\alpha} \times \left\{
\begin{tabular}{cl}
1 & $\nu_{\rm HI} < \nu < \nu_{\rm HeII}$ \\
$f$ & $\nu_{\rm HeII} < \nu$,
\end{tabular}
\right.
\label{eq:jdefn}
\eq
where $J_{{\rm HI},-21}$ is the angle-averaged specific intensity of the background, in units of $10^{-21} \Junits$, $\nu_{\rm HI}$ and $\nu_{\rm HeII}$ are the frequencies corresponding to the \hone and HeII ionization edges, and $f$ is a constant.  We will assume for simplicity that $f=0$ before helium is reionized and $f=1$ afterward.  In our fiducial models, we adopt $\alpha=1.5$ (typical of quasar spectra).  After the quasar that initially ionized a region turns off, the ionizing background  will be provided by stellar sources, and the spectrum will soften considerably.  However, the shape only changes the photoheating rate, which does not substantially affect our calculations.  (It does not affect the HeIII fraction because we set $f=0$ before helium reionization anyway.)

We define the number density of species $i$ to be $n_i \equiv (1+\delta) \tilde{X}_i \bar{\rho}_b/m_p$, where $\bar{\rho}_b$ is the mean proper mass density of baryons.  Note that the $\tilde{X}_i$ are the species number fractions, \emph{not} the neutral fractions.

The thermal evolution of the element is determined by
\bq
{\deriv T \over \deriv t} = -2 H T + {2 T \over 3(1+\delta)} {\deriv \delta \over \deriv t} - {T \over \sum_i \tilde{X}_i} {\deriv (\sum_i \tilde{X}_i) \over \deriv t} + {2 \over 3 k_B n_b} {\deriv Q \over \deriv t},
\label{eq:Tderiv}
\eq
where $\deriv/\deriv t$ is the Lagrangian derivative.  The first term on the right hand side describes the Hubble expansion, the second describes adiabatic cooling or heating from structure formation, the third accounts for the change of internal energy per particle from changing the total particle density, and in the last term $\deriv Q/\deriv t$ is the net heat gain or loss per unit volume from radiation processes (see below).  

The second term requires an expression for the growth of the nonlinear density field.  \citet{hui97} used the Zel'dovich approximation, along with an analytic estimate for the distribution of the strain tensor, which has been shown to provide an accurate analytic approximation to the density evolution in the quasilinear regime.  However, to maintain computational simplicity, we will assume that the gas elements evolve following the spherical collapse (or expansion) model.  We map the linear densities to nonlinear overdensities via the following fitting formula \citep{mo96},
\bq
\delta_L = D(z) \delta_L^0 = \delta_c - {1.35 \over (1+\delta)^{2/3}} - {1.12431 \over \sqrt{1 + \delta}} + {0.78785 \over (1 + \delta)^{0.58661}},
\label{eq:dl-dnl}
\eq
where $D(z)$ is the linear growth function and $\delta_c \approx 1.69$ is the threshold for virialization in the spherical collapse model.  Under this assumption, we can write
\bq
{\deriv \delta \over \deriv t} = \delta_0^L {\deriv D \over \deriv t} {\deriv \delta \over \deriv \delta_L}.
\label{eq:devol}
\eq
In the present context, this approximation affects the temperature distribution (and hence recombination coefficient) 
substantially only at densities far from the mean, so it is not a significant concern for us.

The fourth term includes a number of radiative heating and cooling processes.  The most important heating mechanism is photoionization itself; each species $i$ contributes a term
\bq
\left. {\deriv Q \over \deriv t} \right|_i = n_i \int_{\nu_i}^\infty \deriv \nu \, (4 \pi J_\nu) \sigma_i {{\rm h} \nu - {\rm h} \nu_i \over {\rm h} \nu},
\label{eq:pheat}
\eq
where $\sigma_i$ is the photoionization cross section for species $i$, $\nu_i$ is its ionization threshold, and ${\rm h}$ denotes Planck's constant (to differentiate it from the Hubble constant).  We use the fits of \citet{verner96} for the photoionization cross sections.  The other relevant mechanisms cool the gas, and include Compton cooling off the CMB (which dominates during the era of hydrogen reionization), recombinations (radiative and, for \ion{He}{2}, dielectronic), collisional ionization, collisional line excitation, and free-free emission.  We use the fits of \citet{hui97} for all of these processes except Compton cooling (for which we use the exact form in \citealt{seager99}) and free-free emission (for which we use the fit in \citealt{theuns98}).  We have verified that our results are unchanged if we use the fits presented in \citet{theuns98} (an updated form of those in \citealt{cen92}) for all these mechanisms.

Finally, we have an ionization balance equation for each species; e.g., for hydrogen,
\bq
{\deriv \tilde{X}_{\rm HI} \over \deriv t} = - {\deriv \tilde{X}_{\rm HII} \over \deriv t} =  - \tilde{X}_{\rm HI} \Gamma_{\rm HI} + \alpha_B^{\rm HI}(T) \tilde{X}_e \tilde{X}_{\rm HII} [ \bar{n}_b (1 + \delta) ], 
\label{eq:xievol}
\eq
where $\alpha_B$ is the case-B recombination coefficient and 
\bq
\Gamma_i = \int_{\nu_i}^\infty \deriv \nu \, {4 \pi J_\nu \over {\rm h} \nu} \sigma_i
\label{eq:pion}
\eq
is the ionization rate of species $i$.  We will use $\Gamma_{12} = \Gamma/(10^{-12} \secinv)$ for convenience.  Note that, unlike in \citet{furl07-igmtemp}, we do \emph{not} assume photoionization equilibrium for these calculations.

\section{The Density Distribution} \label{mhr}

Equations~(\ref{eq:Tderiv}), (\ref{eq:devol}), and~(\ref{eq:xievol}) allow us to follow the evolution of a single gas parcel.  We wish, however, to consider the aggregate evolution of large regions of the IGM; we therefore need the distribution of gas density inside the IGM.  Unfortunately, there is no well-tested model at high redshifts.  In the following we will use the density distribution $P_V(\Delta)$ (where the distribution is over volume and $\Delta=1+\delta$) recommended by \citet[henceforth MHR00]{miralda00}, which fits cosmological simulations at $z=2$--$4$ quite well:
\bq
P_V(\Delta) \, \deriv \Delta = A_0 \Delta^{-\beta} \exp \left[ - \frac{(\Delta^{-2/3} - C_0)^2}{2(2 \delta_0/3)^2} \right] \, \deriv \Delta.
\label{eq:pvd}
\eq
Intuitively, the underlying Gaussian density fluctuations are modified through nonlinear void growth and a power law tail at large $\Delta$.  MHR00 argued that the form could be extrapolated to higher redshifts in the following way.  First, $\delta_0$ essentially represents the variance of density fluctuations smoothed on the Jeans scale for an ionized medium (appropriate for our purposes, where we will only consider gas that has been pre-ionized); thus $\delta_0 \propto (1+z)^{-1}$.  The power-law exponent $\beta$ determines the behavior at large densities; we set $\beta=2.5$ for $z>6$.  The remaining constants ($A_0$ and $C_0$) can be set by demanding proper mass and volume normalization.  Note, however, that this distribution comes from a simulation with somewhat different cosmological parameters than the currently preferred values and so is probably not accurate enough for detailed comparisons.

MHR00 also offer a prescription for determining $\lambda_i$, the mean free path of ionizing photons, in a highly-ionized universe.  They assume that the gas density field has two phases:  low-density gas is highly ionized and transparent to ionizing photons, while high density regions (with $\Delta>\Delta_i$) are neutral because of self-shielding.  We will examine the accuracy of this assumption later, but it will be quite useful for most of our work (see also \citealt{furl05-rec}).  In this picture, the mean free path equals the mean distance between the self-shielded clumps along a random line of sight, which is approximately
\bq
\lambda_i = \lambda_0 [1 - F_V(\Delta_i)]^{-2/3}.
\label{eq:mfp-mhr}
\eq
Here $F_V(\Delta_i)$ is the fraction of volume with $\Delta < \Delta_i$ and $\lambda_0$ is a (redshift-dependent) normalization factor.  Formally, this expression is valid only if the number density and shape (though not total cross section) of absorbers is independent of $\Delta_i$.  This is obviously not true in detail for the cosmic web.  However, MHR00 found that equation~(\ref{eq:mfp-mhr}) provides a good fit to numerical simulations at $z=2$--$4$ with $\lambda_0 H(z) = 60 \kms$ (in physical units).  We will extrapolate the same prescription to higher redshifts.  Equation (\ref{eq:mfp-mhr}) was derived in the highly ionized limit,
ignoring photoelectric absorption by low-column density systems, and
therefore generically overestimates $\lambda_i$ at the ionization
edge.  The correction can be up to a factor of $\sim 2$ even in 
highly ionized cases \citep{furl05-rec}.  Some of this decrease can be
compensated by the longer mean free paths of higher-energy photons,
but to be conservative, we set $\lambda_0 H(z) = 30 \kms$ whenever we
refer to the MHR00 mean free path in our calculations. We will discuss
the accuracy of this mean free path estimate in more detail below.

The MHR00 prescription was developed for the post-overlap universe, when dense regions are self-shielded and the mean-free path is controlled by Lyman-limit systems. In the pre-overlap universe, assuming that all regions with $\Delta < \Delta_{i}$ are ionized implicitly assumes an ``outside-in" reionization topology, in apparent contradiction with expectations from semi-analytic models and simulations (e.g., \citealt{furl04-bub,mcquinn07}) that reionization is an ``inside-out" process due to source bias. The MHR00 prescription is likely applicable during reionization once the mean free path is sufficiently large that a typical photon samples a fair (and relatively unbiased) fraction of the density distribution during its flight \citep{zhang07}.\footnote{Note that the details of the small-scale density distribution within and around halos are generally absorbed into the escape fraction.} The MHR00 prescription is particularly well suited to quasar fossil bubbles.  For one, the {\it initial} mean free path within the quasar bubble (tens of Mpc) is comparable to that of the post-overlap universe, when $F_{V}$ is close to unity; one is considering a sufficiently large volume that it is a fair sample of the density distribution. \citet{lidz07} find that the bias is modest when averaged over the quasar bubble volume.  Moreover, dense regions recombine first, leaving underdense regions ionized---precisely the assumptions made in the MHR00 model, provided the ionizing background is relatively uniform (see \S \ref{typical} for more discussion of the latter point). 

\begin{figure}
\plotone{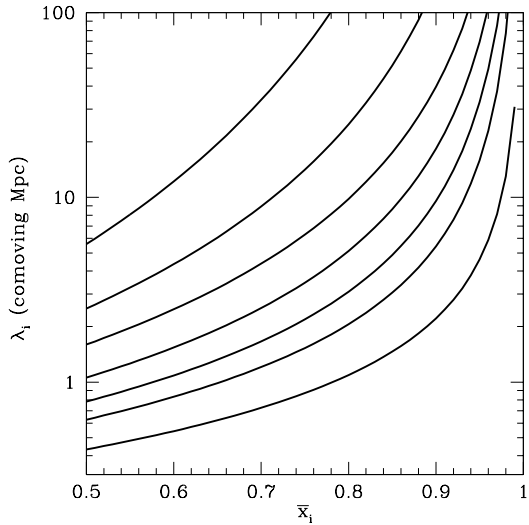}
\caption{Comoving mean free path of photons at the hydrogen ionization edge as a function of the ionized fraction $\bxion$ in the MHR00 model.  The curves assume $z=2,\,3,\,4,\,6,\,8,\,10$, and $15$, from top to bottom.}
\label{fig:mhrmfp}
\end{figure}

Figure~\ref{fig:mhrmfp} shows $\lambda_i$ as a function of mass-averaged ionized fraction $\bar{x}_{i} \equiv F_{M}(\Delta_{i})$ for several different redshifts ($z=2,\,3,\,4,\,6,\,8,\,10$, and $15$, from top to bottom).  Note that $\lambda_i$ decreases as redshift increases because the IGM gets less and less clumpy, so the spacing between gas above a given density threshold decreases.  At higher redshifts, $\bxion \approx 0.5$ corresponds to $\Delta \approx 1$, so these regions are typically separated by a fraction of a Mpc.

It is important to note that the MHR00 mean free path is \emph{much} larger than one would expect in a uniformly ionized medium.  The optical depth in a uniform medium, measured across one fiducial MHR00 mean free path, can be expressed as
\bqa
\tau_{\lambda} & \sim & \xhi \bar{n}_H \lambda_i \sigma_{\rm HI} \label{eq:tau_simple} \\
& \sim & 260 \ \xhi \left( {\lambda_i \over 3 \Mpc} \right) \left( {1+z \over 9} \right)^2 \nonumber
\eqa
where $\xhi$ is the mean neutral fraction of the gas \emph{outside} of self-shielded regions, $\bar{n}_H$ is the mean density of hydrogen nuclei, and $\sigma_{\rm HI}$ is the photoionization cross section.  For the latter, we used $\sigma_{\rm HI}=\bar{\sigma}=2 \times 10^{-18} \cmsq$, the value averaged over  a typical stellar spectrum \citep{miralda03}.  Figure~\ref{fig:mhrmfp} shows that under the MHR00 geometry,
the mean free path reaches $\sim$Mpc scales when  $\xhi \sim 0.1-0.5$.
In comparison, in a uniform IGM, reaching this large mean free path would require $\xhi \la 10^{-3}$.
The reason is that the MHR00 procedure calculates the mean free path in an entirely different way:  it \emph{assumes} that the bulk of the IGM is highly-ionized (with negligible absorption) except for regions above the self-shielding threshold (which are entirely neutral).  In other words, in the MHR00 geometry, $\bxion$ (which is a mass average) refers to the fraction of gas outside these self-shielded regions, not to the volume-averaged ionized fraction.  We will check for the consistency of this picture later and see some examples in which it becomes untenable.

\section{Fossil Bubbles with Zero Ionizing Background} \label{zeroion}

We begin by considering the simple case of an ionized bubble whose generating source turns off instantaneously, with no sources remaining or turning on.  Thus the gas simply recombines, without any further influence from a radiation field. In the context of hydrogen reionization, pre--existing galaxies in the fossil bubble will always maintain an ionizing background. Considering the zero--background case is nevertheless a useful pedagogical exercise; furthermore, in the context of helium reionization, where the death of the quasar may leave no ionizing sources, this case is also physically relevant.

To produce our ionization histories, we initialize our calculation at an assumed redshift $z_i$ where the quasar has \emph{fully} ionized a region.  We ignore the residual neutral gas at this stage, including any self-shielded, high--density clumps that may remain neutral even in the presence of a quasar.  We then compute $\xion (\Delta,z)$ at a final redshift $z$.  The mass-averaged ionized fraction within the bubble, $\bar{x}_{i,m}$, is
\bq
\bar{x}_{i,m} = f_{\rm IGM}^{-1} \int_0^{\Delta_{\rm max}} \deriv \Delta \, \xion (\Delta,z) \Delta P_V(\Delta).
\label{eq:ximass}
\eq
We set $\Delta_{\rm max}=50$ so as to exclude virialized gas from the calculation (since its ionization properties are irrelevant to the IGM properties).  The prefactor $f_{\rm IGM}^{-1}$ is the fraction of mass in gas with $\Delta<\Delta_{\rm max}$ and normalizes the result to give the ionized fraction of just the IGM gas.  The volume-averaged ionized fraction, $\bar{x}_{i,v}$ is computed similarly, except with $\Delta P_V(\Delta) \rightarrow P_V(\Delta)$ inside the integrand and $f_{\rm IGM}$ the volume fraction of gas with $\Delta<\Delta_{\rm max}$.

\begin{figure}
\plotone{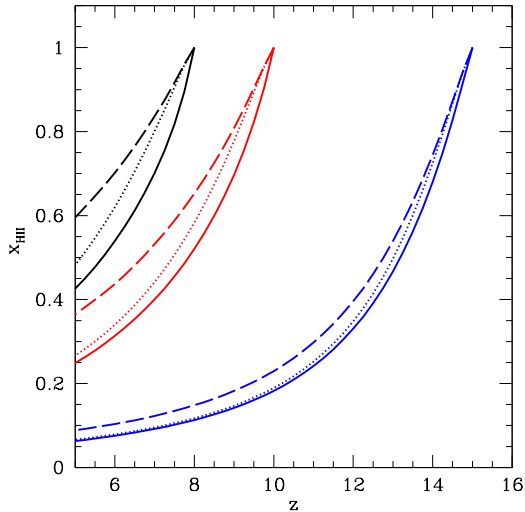}
\caption{Ionization histories for fossil HII regions that are produced when the ionizing source turns off at some initial redshift $z_i$.  The fossils are assumed to see no ionizing flux, so the gas is allowed to recombine.  The three sets of curves assume $\xion=1$ at $z_i=8,\,10,$ and $15$, from left to right.  Within each set, the solid and dashed curves show the mass-- and volume--averaged ionized fractions, $\bar{x}_{i,m}$ and $\bar{x}_{i,v}$, respectively, using the MHR00 fit for the density distribution.  The dotted curves show the histories for elements at the mean density (with $\Delta=1$).}
\label{fig:fossil}
\end{figure}

Figure~\ref{fig:fossil} shows the evolution of the ionized fraction in a fossil bubble with $\Gamma_{12}=0$.  We consider three initialization redshifts, $z_i=8,\, 10,$ and $15$.  For each case, we show $\bar{x}_{i,m}$, $\bar{x}_{i,v}$, and $x_i(\Delta=1)$ by the solid, dashed, and dotted curves (respectively).  As one would expect from the short recombination time at these high redshifts, all of these cases are able to recombine significantly. Also as one would expect, $\bar{x}_{i,m} < \bar{x}_{i,v}$, because underdense regions (which recombine slowly) of course fill more than their share of the volume.  However, the two are not so far apart, and in particular the mass-averaged value is always quite close to the expectation for gas elements at the mean density.  

\begin{figure}
\plotone{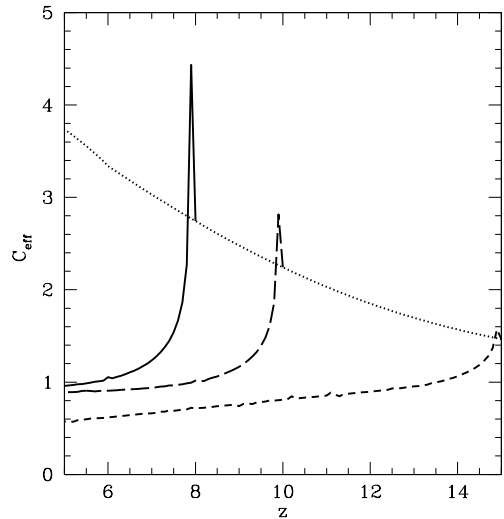}
\caption{The effective clumping factors for the recombining fossils in Fig.~\ref{fig:fossil}.  For reference, the dotted curve shows $C$ for the MHR00 IGM model, assuming that all gas at $\Delta<50$ is fully ionized (with gas above this threshold residing inside collapsed halos and excised from the IGM).  }
\label{fig:clump}
\end{figure}

Many models (including both semi-analytic and numerical varieties) compute reionization histories using a so-called clumping factor, $C \equiv \VEV{n_e^2}/\VEV{n_e}^2$ to estimate the total recombination rate.  The dotted curve in Figure~\ref{fig:clump} shows this factor in the MHR00 model as a function of redshift, assuming that all gas with $\Delta<50$ is ionized.  The effective value is always larger than unity, but not by an extremely large value; it is significantly smaller than often assumed because we have excluded virialized systems and also assumed that the gas is Jeans-smoothed on the scale appropriate for ionized gas.  It increases toward lower redshifts as structure formation continues and amplifies density contrasts.

Unfortunately, blindly applying this clumping simplification only works when \emph{all} the gas is in ionization equilibrium.  In reality, this is never a particularly good assumption.  For example, during reionization, the dense interiors of self-shielded clumps remain neutral, hardly ever see ionizing photons, and do not contribute to the ionized volume.  Ionizing photons need not battle recombinations in this gas, but only the slower recombinations in the more rarefied outskirts of such regions.  The dotted curve essentially deals with this problem by excluding dense gas from the clumping calculation (as is typical in numerical simulations; e.g., \citealt{iliev05-sim, mcquinn07}).  In reality, one should allow the maximum density threshold $\Delta_i$ to evolve to reflect the actual ionizing background (and neutral fraction) within the bubble.

The problem is somewhat more insidious for fossil bubbles.  The solid, long-dashed, and short-dashed curves in Figure~\ref{fig:clump} show the effective clumping factors for fossil bubbles, again with initial ionization at $z_i=8,\,10,$ and $15$.  Here we define the clumping factor $C_{\rm eff}$ by comparing the overall recombination rate to that for a gas element at the mean density.  After an initial spike caused by rapid cooling of the dense gas (remember $\alpha \propto T^{-0.7}$),\footnote{This spike is probably not physical, because the gas is actually heated over the lifetime of the quasar and expands and cools continuously, rather than being set to an arbitrary initial temperature; see \citet{shapiro04,mesinger06} for discussions of the thermal behavior of photoheated gas in simulations.} $C_{\rm eff}$ quickly decreases to a value close to unity.

This behavior occurs because there is (by definition) zero ionizing background inside fossils, so \emph{no} gas elements can be in ionization equilibrium.  Although dense elements recombine quickly, they cannot help their less-dense neighbors to do so; once a region is mostly neutral, it breaks off from the recombining gas.  Thus the density of an ``average'' recombining parcel decreases with cosmic time, and with it the effective clumping factor.  It eventually levels off at $C_{\rm eff} \la 1$, because the recombination time for gas at the mean density is comparable to the Hubble time.  This behavior is vital for analytic models of fossil bubbles -- and also for any model with a ``recombination era'', including so-called ``double reionization'' -- and a proper treatment requires a full accounting of the IGM density distribution.  This also implies that cosmological simulations cannot use simple subgrid clumping in any cells where the local ionizing background is zero; instead the instantaneous recombination rate depends on the integrated recombination history of the cell.

\begin{figure}
\plotone{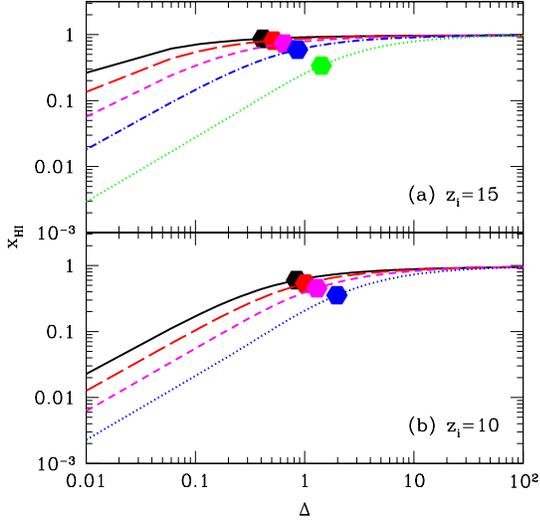}
\caption{Neutral fraction as a function of density inside fossil HII regions.  \emph{(a):} $z_i=15$.  The curves show $z=14$--$6$, from bottom to top, with $\Delta z = 2$; these have $\bar{x}_{{\rm HI},m}=0.32,\,0.67,\,0.82,\,0.89,$ and $0.93$.  \emph{(b):} $z_i=10$.  The curves show $z=9$--$6$, from bottom to top, with $\Delta z = 1$; these have $\bar{x}_{{\rm HI},m}=0.68,\,0.5,\,0.62$, and $0.70$.  The filled hexagons mark the nominal $\Delta_i$ values, which, in the MHR00
 model, produce the same average neutral fraction.}
\label{fig:xi_Delta}
\end{figure}

Another way to see this is through the neutral fraction as a function of density.  Figure~\ref{fig:xi_Delta} shows sequences of $\xhi(\Delta)$ at several different times, with $z_i=15$ and $z_i=10$ in the top and bottom panels, respectively.  For the most part, low-density gas remains highly-ionized, while dense regions quickly recombine.  However, the transition between the two is rather extended (because the recombination time is proportional to $\Delta^{-1}$).  Indeed, even deep voids can have a substantial neutral fraction long after the quasar turns off.

\subsection{The Mean Free Path}
\label{mfp-noion}

Next we will consider how the MHR00 model for the ionized gas distribution and mean free path fares in this scenario.  Qualitatively, one might expect it to perform adequately:  after all, the distribution goes from relatively highly-ionized gas when $\Delta \ll 1$ to nearly neutral gas at high densities.  In fact, blindly using the ionized fraction to fix $\Delta_i$ (employing the physical scenario of MHR00, and assuming that gas with $\Delta < \Delta_i$ is actually highly ionized) even roughly reproduces the regime for which $x_i \sim 0.5$ (see Fig.~\ref{fig:xi_Delta}).  \emph{However}, $\Delta_i$ clearly does not mark a sharp separation between highly-ionized and nearly neutral gas; rather, the low-density gas is also marginally neutral.  As a result, the mean free path for these fossils will be far smaller than shown in Figure~\ref{fig:mhrmfp}.  Simply assuming a uniform medium (eq.~\ref{eq:tau_simple}) yields $\tau \sim 1$ across a region of size $\lambda_i \sim 7 \kpc$ at $\xhi \sim 0.5$ -- no larger than a typical halo!

We can do slightly better than this uniform IGM approximation by dividing the gas into discrete systems in a similar way to the \lya forest at lower redshifts \citep{schaye01}.  The typical length scale of such a system is the Jeans length, $L_J \sim G/\sqrt{c_s \rho}$, where $c_s$ is the local sound speed.  Then the column density of neutral hydrogen (which determines the optical depth of each system to ionizing photons) is $N_{\rm HI} \sim n_H L_J \propto n_H^{1/2} T^{1/2}$.  At the Lyman edge,
\bq
\tau \sim 10 \left( {\Delta \over 0.1} \, {\xhi \over 0.1} \right)^{1/2} \left( {1+z \over 8} \right)^{3/2}.
\label{eq:tau_sys}
\eq
Comparison to Figure~\ref{fig:xi_Delta} shows that the neutral fractions required to produce $\tau>1$ are typically exceeded even in low-density gas.  Thus, effectively, \emph{every} discrete absorber -- even a deep void -- is a Lyman-limit system, and the mean free path of an ionizing photon is extremely small.  The MHR00 model is obviously not applicable in this scenario.

\section{A Nonzero Ionizing Background} \label{nzion}

We now include a small, but nonzero, ionizing background in our calculations. 
 This background could represent pre--existing galaxies, as well as any new galaxies that continue to form within the fossil bubble, and/or any low-level residual emission from the quasar after it has turned off (see \S \ref{post-quasar} below).  To set the amplitude during hydrogen reionization, we begin by assuming that the emissivity is simply proportional to the rate at which gas collapses into star-forming halos, $\deriv \fcoll/ \deriv t$. Here $\fcoll$ is the collapsed fraction in halos above the minimum mass $\mmin$ able to form stars. We will parameterize this threshold mass in terms of the mass $m_4$ of halos with virial temperatures larger than $10^4 \kel$, so that atomic cooling is efficient \citep{barkana01}.\footnote{Note that it is possible for the effective $\mmin$ to differ between the universe at large and the highly-ionized fossil bubble, because (for example) photoheating may increase the Jeans mass \citep{rees86, efstathiou92, thoul96, dijkstra04-feed}.}    We assume an ionizing efficiency $\zeta$, such that the globally-averaged ionized fraction $\bxion^g$ at the initial redshift $z_i$ is\footnote{Note that this expression ignores recombinations, but in our calculations they can be simply swept into the free parameters described below, so long as they are roughly uniform.}
\bq
\bxion^g(z_i) = \zeta \fcoll(z_i).
\label{eq:zetadefn}
\eq

\subsection{Gas at the Mean Density} \label{meanden}

We can now take two possible approaches.  In this section, we will discuss an approach similar to \S~\ref{zeroion}, i.e. tracking the evolution of each gas element independently, simply adding a redshift-dependent ionizing background to the evolution equations.  This approach is imperfect, as we will discuss below, but it is useful to provide a feel for the effectiveness of an ionizing background in suppressing recombinations.  In the next section, we will follow an alternative (more global) approach, based directly on the emissivities (rather than fluxes).

We begin by examining the evolution of gas with $\Delta=1$. The ionization rate can be written as
\bqa
\Gamma & = & \int d\nu \, \epsilon_\nu \lambda_\nu \sigma_{\rm HI}(\nu)
\label{eq:gamma-emiss} \\
& = & \bar{n}_b \bar{\sigma} \bar{\lambda} { \deriv (\zeta \fcoll) \over \deriv t},
\label{eq:gamma-fcoll}
\eqa
where $\epsilon_\nu$ is the emissivity per unit frequency and $\lambda_\nu$ is the (proper) mean free path of a photon with frequency $\nu$.  Substituting values appropriate for our cosmology, we find
\bq
\Gamma_{12} = 0.018 \secinv {\bxion^g(z_i) \over \fcoll(z_i)} \left( {\lambda_i \over {\rm Mpc}} \right) \left( {1+z \over 8} \right)^{9/2} \left| {\deriv \fcoll \over \deriv z} \right|,
\label{eq:gamma-num}
\eq
where $\lambda_i$ is the comoving mean free path (averaged over frequency) and we have used equation~(\ref{eq:zetadefn}) to eliminate the efficiency $\zeta$ in terms of the more physically transparent globally-averaged ionized fraction when the bubble is first ionized.\footnote{In fact, bright quasars form in the most massive halos and hence in biased regions of the IGM, inside of which $\fcoll$ should be larger than its global mean.  \citet{lidz07} have shown, however, that although the bias is large within a few Mpc of the quasar, it is relatively modest when averaged over the entire volume of a large fossil bubble, so we ignore it for simplicity.}

\begin{figure}
\plotone{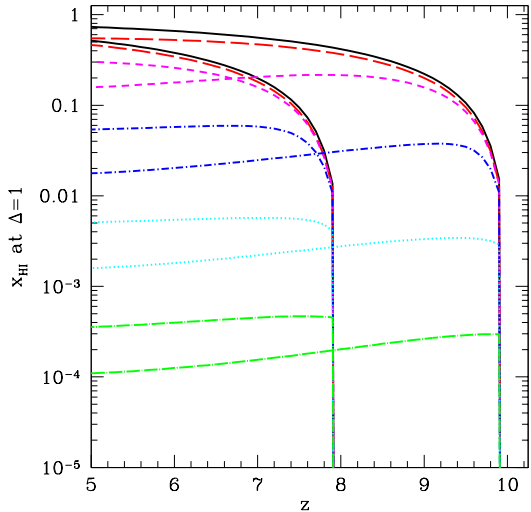}
\caption{Ionization histories for gas with $\Delta=1$ under a uniform ionizing background.  The two sets of curves assume $\xion=1$ at $z_i=8$ and $10$.  The solid curves assume $\Gamma=0$ as in Fig.~\ref{fig:fossil}.  The long-dashed, short-dashed, short dot-dashed, dotted, and long dash-dotted curves within each set (i.e., from top to bottom) take $\bxion^{g}(z_i) \lambda_i = 0.6, \, 6,\, 60,\,600$ and $6000 \kpc$, respectively.  Note that this calculation breaks down when self-shielding becomes important; see text.}
\label{fig:mfp}
\end{figure}

Figure~\ref{fig:mfp} shows ionization histories for gas at the mean density, initially ionized at $z_i=8$ and $10$. The solid curves assume $\Gamma=0$.  The long-dashed, short-dashed, short dot-dashed, dotted, and long dash-dotted curves within each set (from top to bottom) take $\bxion^{g}(z_i) \lambda_i = 0.6, \, 6,\, 60,\,600$ and $6000 \kpc$, respectively. This combination of parameters sets the overall amplitude of the ionizing background in equation~(\ref{eq:gamma-num}); the background can be large either because one begins late in reionization ($\bxion^g$ is large) or because the mean free path in the fossil bubble is large.  For comparison to results below, it is useful to think of these as having $\bxion^{g}(z_i)=0.2$, so that the bubble is created early in reionization (but not too early) and $\lambda_i=0.003$--$30 \Mpc$.  Comparison to Figure~\ref{fig:mhrmfp} shows that the MHR00 model could apply to the IGM in the latter two cases (3-30 Mpc); the other cases imply very small mean free paths and correspond to a regime where the MHR00 assumption breaks down.

\subsection{The Mean Free Path}
\label{mfp-uniform}

We will next explicitly consider how the MHR00 model for the mean free path fares in this picture.  We must immediately note that Figure~\ref{fig:mfp} shows the evolution only for gas at the mean density.  However, in the interesting limit in which this gas remains highly ionized, the fossils approach an asymptotic ionized fraction relatively quickly:  this is the value appropriate for photoionization equilibrium, which states
\bqa
{x_{\rm HI} \over (1 - x_{\rm HI})^2} & = & 1.5 \times 10^{-3} \Delta \left[ {\bxion^g(z_i) \lambda_i \over {\rm Mpc}} \right]^{-1} \left( {1+z \over 8} \right)^{-3/2} \nonumber \\
& & \times \left[ {\fcoll(z_i)  \over \deriv \fcoll / \deriv z} \right]^{-1}.
\label{eq:ioneq}
\eqa
Here we have assumed $T=10^4 \kel$; in fact, the temperature evolves as the universe expands, so this expression is only approximately correct.  Except for the brief initial adjustment phase, the equilibrium value works extremely well for $\bxion^{g}(z_i) \lambda_i \ga 60 \kpc$ (so long as the appropriate temperature is chosen).  Thus we expect $x_{\rm HI} \propto \Delta$ in other gas parcels, so long as the ionizing background illuminates every region equally.

Now let us suppose that a quasar appears at $z_i$, ionizes a large bubble (with comoving radius $\ga 30 \Mpc$), and turns off.  For a concrete example, let us assume that $\bxion^g(z_i)=0.2$.  The fossil will immediately begin recombining.  At this initial stage, the mean free path must be at least as large as the quasar bubble (otherwise a single quasar could not have ionized the entire region).  Thus the bottom curve in Figure~\ref{fig:mfp} will be most appropriate, and the neutral fraction will initially be $\sim 3 \times 10^{-4}$.  Substituting into equation~(\ref{eq:tau_simple}), the optical depth across one mean free path is $\sim 0.8$.  Thus the mean free path would remain large, and the entire fossil would clearly remain highly ionized.

Such long mean free paths are quite reasonable at high redshifts, even outside of fossils.  For example, \citet{lidz07} show that extrapolations of the \lya forest to $z \ga 6$ yield mean free paths at the ionization edge of $\sim 8$--$15 \Mpc$.   Moreover, as described above, the MHR00 model predicts $\lambda_i \ga 10 \Mpc$ for the highly ionized gas immediately after the quasar turns off.  

With the help of equation~(\ref{eq:ioneq}), we can also see that denser gas will also remain quite ionized: in the example above, $\Delta=50$ gas would initially have $\xhi \sim 0.015$.  However, such an approach ignores radiative transfer, which is crucial for our purposes (it is instead similar to the way many cosmological simulations mimic a uniform radiation field by illuminating each and every gas particle identically).  The basic point is that dense gas elements are not isolated:  rather they are part of discrete regions, surrounded by other dense elements.  The accumulated opacity of the outskirts of such objects reduces the effective ionizing background in the densest elements -- or in other words they become self-shielded.  When this process is included, it is possible to approximate the IGM as two-phase, with gas at $\Delta < \Delta_i$ highly ionized (and eventually settling into ionization equilibrium) and gas at $\Delta > \Delta_i$ self-shielded and fully neutral (MHR00, \citealt{furl05-rec}).  It is these self-shielded elements that set the mean free path in the MHR00 picture, and it is also these regions that the simple-minded approach of this section ignores.  

\section{Fossil Bubbles and Inhomogeneous Reionization} \label{inhom}

To rectify the shortcomings of the previous section, we note that the principal problem is that a uniform ionizing background allows too much ionization.  Because each gas parcel is treated independently, and because the ionizing background is not re-adjusted to account for losses due to recombinations, photons are not conserved in such a scheme.  In more concrete terms, the scheme above not only puts photons in the outskirts of dense regions but also in their centers; self-shielding demands that the outskirts consume all the photons, so adding more to the center must ``over-ionize'' the universe.

We will therefore follow a more global approach to the evolution of the ionized fraction by enforcing photon conservation as our starting point.  We first ask what emissivity is required to overcome the recombinations within an ionized bubble.   As in equation~(\ref{eq:zetadefn}), we parameterize the emissivity in terms of the equivalent global ionized fraction outside the fossil.

\begin{figure}
\plotone{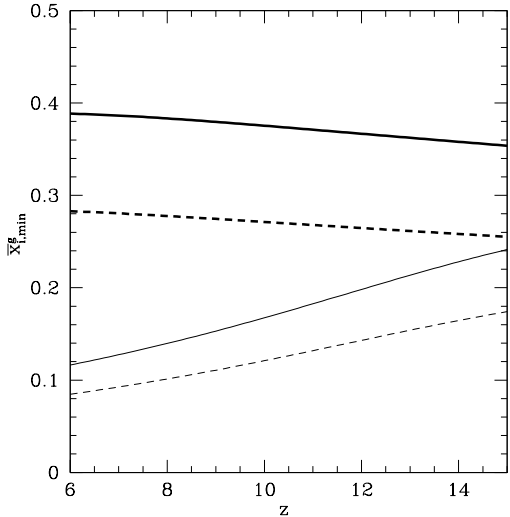}
\caption{Minimum global ionized fraction required for the background ionizing sources to counteract recombinations in fossil bubbles.  The thick curves use the MHR00 clumping factor, while the thin curves assume $C_{\rm eff}=1$.  The solid curves assume $\mmin = m_4$, while the dashed curves assume $\mmin = 10 m_4$. The figure suggests that fossils can be kept highly ionized, at a level well above the ionized fraction in the rest of the IGM.}
\label{fig:xi_min}
\end{figure}

Figure~\ref{fig:xi_min} shows $\bar{x}_{i,min}^g$, the minimum $\bxion^g$ required to overcome recombinations in a \emph{fully-ionized} bubble (in this context, ``fully-ionized" implies that all gas elements below $\Delta_{i}=50$ are ionized).  Note that if the emissivity was constant at all times, then
$\bar{x}_{i,min}^g$ would simply equal $t_{\rm gal}/t_{\rm rec}$, the ratio of
the time elapsed since the onset of galaxy formation to the instantaneous
recombination time.  In practice, the emissivity increases steeply
with cosmic time, which decreases the value of $\bar{x}_{i,min}^g$.\footnote{Unlike in the previous section, here we take the global ionized fraction at each redshift, not at the moment of the bubble's creation.  Also note that our conversion from emissivity to $\bxion^g$ ignores recombinations in the background medium, so in reality the required value will be even smaller.} Note further that the recombination rate is proportional to the ionized fraction, so the minimum requirement for a particular bubble should actually be multiplied by its mean ionized fraction at the appropriate redshift.  We show results for two clumping factors: the upper thick set of curves take the MHR00 value and are appropriate for the period shortly after the fossil forms, while the thin lower curves assume $C=1$.  The latter is appropriate for evaluating whether relatively old fossil bubbles that have partially recombined will continue to do so once the dense gas is already neutral (see Fig.~\ref{fig:clump}).  The solid curves assume $\mmin=m_4$, while the dashed curves assume $\mmin=10m_4$.  The latter, higher mass threshold presents an even weaker requirement on $\bxion^g$, because more massive halos evolve more quickly over cosmic time, so that the emissivity that can keep the fossils ionized corresponds to an even lower background ionized fraction.

We see that, even taking a conservative clumping factor, fossil bubbles can only recombine when $\bxion^g \la 0.3$--$0.4$.  With the MHR clumping value, the minimum increases slowly as redshift decreases as the IGM becomes clumpier.  If, on the other hand, we look at the later stages when $C \sim 1$ (or roughly $\bxion \la 0.75$), the requirement decreases to $\bxion^g \ga 0.1$--$0.2$.  In the latter case, the minimum decreases as redshift decreases because the mean IGM density (and hence recombination rate) falls as well.  This simple calculation suggests that quasar bubbles will only recombine significantly if they are created well before reionization is underway, at $\bxion^g \la 0.1$--$0.3$.

To give some concrete scenarios, we track the ionized fraction within the fossil via
\bq
{\deriv \bxion \over \deriv z} = \zeta {\deriv \fcoll \over \deriv z} - \bxion C \alpha_A \bar{n}_e {\deriv t \over \deriv z},
\label{eq:dxiondz}
\eq
where $\bar{n}_e$ and $\bar{n}_p$ are the the  mean electron and proton densities at redshift $z$.  We assume (unless stated otherwise) that $\mmin=m_4$ to calculate $\fcoll$.  For the effective clumping factor, we will generally use the MHR00 value $C_{\rm MHR}$ but also compare to a case with $C=1$ (where the recombination rate is slower).  This assumes that the same stellar sources that set $\bxion^g(z_i)$ continue to illuminate the fossil.  Of course, if the fossil persists for long enough, and if quasars are relatively common, it is possible that another quasar will form in the same region; in that case, the fossil will quickly become completely ionized again.

\begin{figure*}
\plottwo{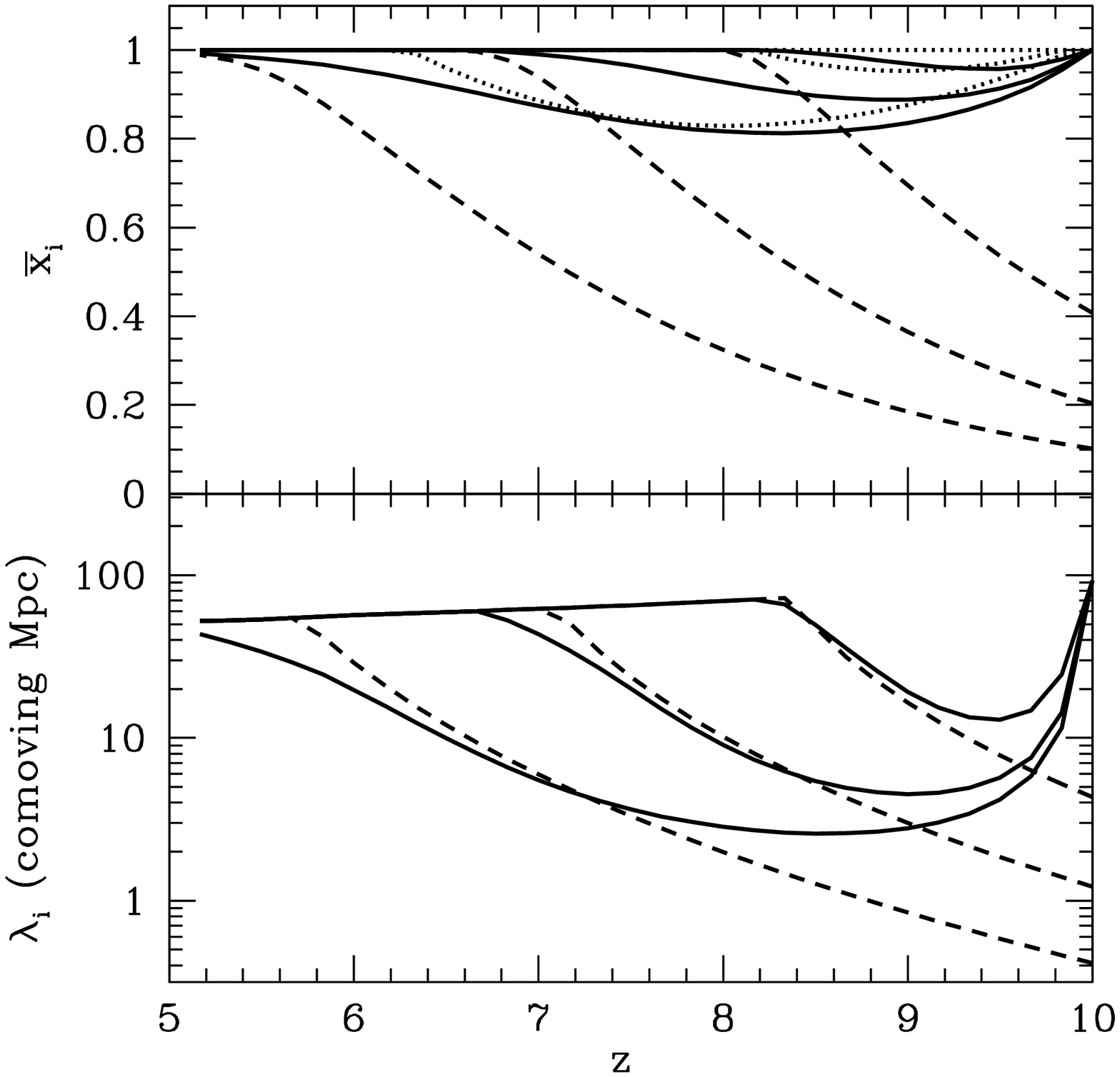}{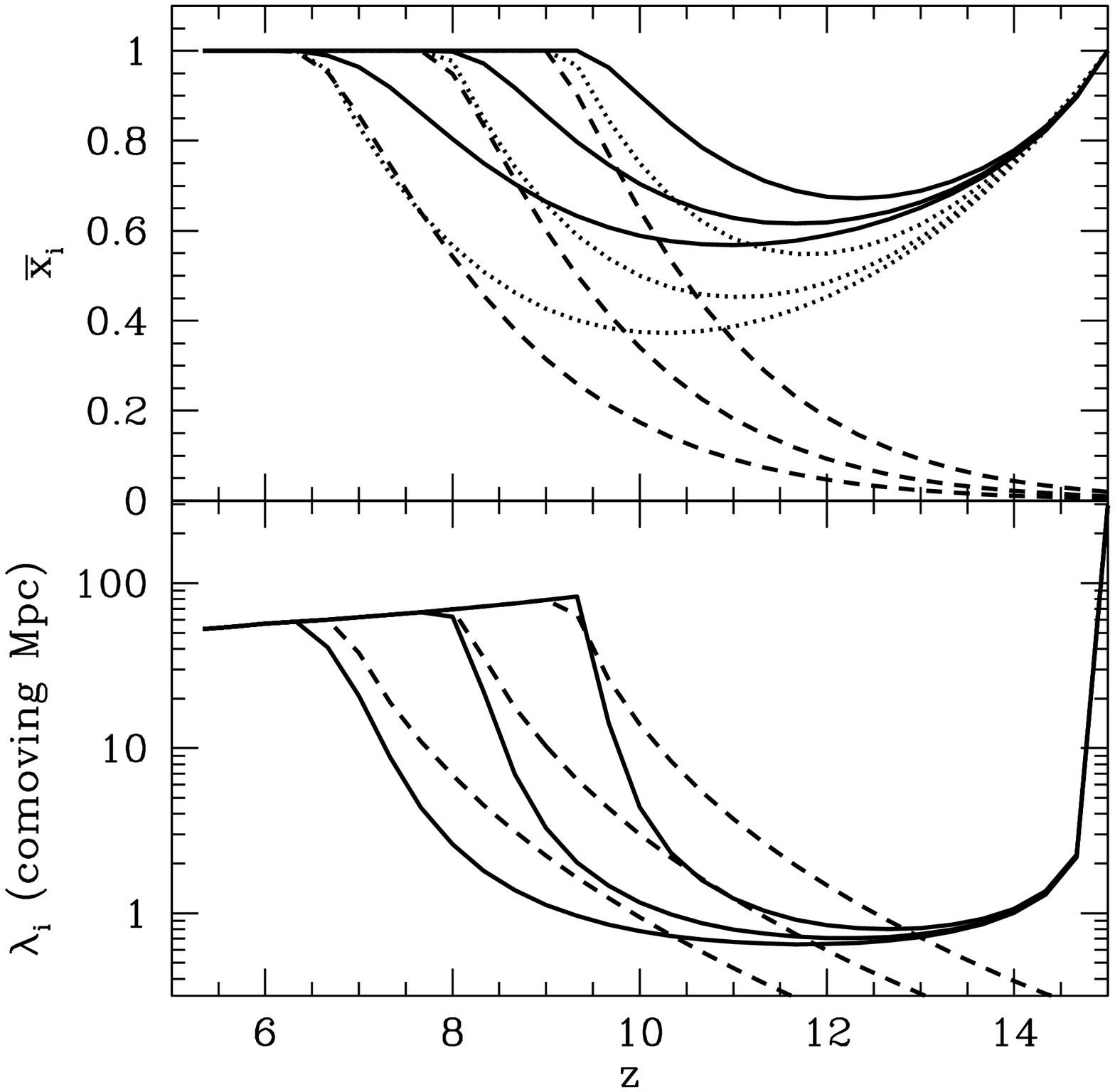}
\caption{\emph{Top:}  Comparison of the ionization histories for fossil bubbles and the universe as a whole.  In each panel, the solid and dotted curves are for fossil bubbles with $C=C_{\rm MHR}$ and $C=1$, respectively.  The dashed curves show $\bxion^g$ (with $C=C_{\rm MHR}$).  \emph{Left:}  $z_i=10$, with the three curves in each set taking $\bxion^g(z_i)=0.4,\,0.2,$ and $0.1$, from top to bottom.  \emph{Right:}  $z_i=15$, with the three curves in each set taking $\bxion^g(z_i)=0.02,\,0.01,$ and $0.005$, from top to bottom. \emph{Bottom:}  Mean free paths of ionizing photons in the fossil and in the background universe.  The solid curves show $\lambda_i$ from the MHR00 model; the dashed curves show a fiducial clustering length $R_c$ for galaxies according to $\bxion^g(z)$ (see text for discussion).  The parameters are the same as in the top panel, except that we do not show the cases with $C=1$.}
\label{fig:xicomp}
\end{figure*}

Figure~\ref{fig:xicomp} shows the resulting ionization histories of the fossil bubbles, in comparison to those of the background, $\bxion^g(z)$, for a series of assumed emissivities (for the latter, we can also use eq.~\ref{eq:dxiondz}, but with different initial conditions; we use $C_{\rm MHR}$ for the background, but it makes little difference).  In each panel, the solid and dotted curves are for fossil bubbles with $C=C_{\rm MHR}$ and $C=1$, respectively.  The dashed curves show $\bxion^g$.  In the left panel, $z_i=10$, with the three curves in each set taking $\bxion^g(z_i)=0.4,\,0.2,$ and $0.1$, from top to bottom.  In the right panel, $z_i=15$, with the three curves in each set taking $\bxion^g(z_i)=0.02,\,0.01,$ and $0.005$, from top to bottom. We now discuss each of these cases in turn.

\subsection{A Typical Quasar} \label{typical} 

As seen in Figure~\ref{fig:xicomp}, the requirement that reionization completes by $z=6$ forces $\bxion^g(z_i) \ga 0.15$ for $z_{i} \le 10$. So, for quasars at $z_i=10$, the contrast between the background and the fossils (or, indeed, between the background and active quasar bubbles) is never more than a factor of a few.  Unless supermassive black holes form long before reionization, this case is therefore typical of most fossils.  Figure~\ref{fig:xicomp} shows that, as expected, the ionizing background inside the bubble efficiently suppresses recombinations -- in no case does $\bxion$ fall below $\sim 0.8$ for these fossils.  This is true regardless of the clumping prescription -- while the MHR00 model allows a more rapid initial decline, the two prescriptions converge once the dense gas recombines and the clumping factor approaches unity.  

Our description is still, however, incomplete, because we have not answered one important question:  as they recombine, do the fossils re-develop the ``swiss cheese'' topology typical of reionization, with islands of ionized gas separated by a sea of neutral (or in this case partially neutral) hydrogen? Up to this point, we have assumed that ``shadowing'' and other radiative transfer effects are unimportant, so that photons are delivered to any region that is not self-shielded as needed.  Of course, in reality the ionizing sources are clustered, which modulates the local emissivity.  During reionization of the background universe, this strongly affects the distribution of bubble sizes \citep{furl04-bub}.  Is the same true for recombining fossils?

To answer this, we will need to compare the mean free path within the fossil to the typical scale of fluctuations in the ionizing background generated by the clustering of the ionizing sources.  First we consider the fossils.  Our primary tool for computing $\lambda_i$ is the MHR00 model.  Given an ionized fraction, this sets a threshold density $\Delta_i$ above which gas is assumed to be self-shielded, and below which it is highly ionized.  In our case, $\Delta_i$ can be determined by demanding that the fraction of mass with $\Delta<\Delta_i$ is equal to $\bxion$; with that, $\lambda_i$ follows from equation~(\ref{eq:mfp-mhr}).

The solid lines in the bottom panels of Figure~\ref{fig:xicomp} show this prescription; they correspond to the same models as in the upper panels, all using $C_{\rm MHR}$ for the clumping.\footnote{Note that we use $\lambda_0=30 \kms$ here, as described in \S \ref{mhr}.  This is a conservative \emph{lower} bound on the mean free path, because of high-energy photons, so if anything we overestimate the probability that regions break off from the uniform background.}  Obviously, the mean free path falls immediately after the bright quasar turns off;\footnote{In actuality, the initial mean free path is probably no larger than the fossil, so the decline may not be quite so severe except for the brightest quasars.} it reaches a minimum simultaneously with $\bxion$ and then increases again.  In our models, we assume that all gas with $\Delta>50$ remains neutral, so the mean free paths eventually match onto $\lambda_i$ corresponding to this density threshold (the turnover in the solid curves when $\bxion \approx 1$).

\begin{figure}
\plotone{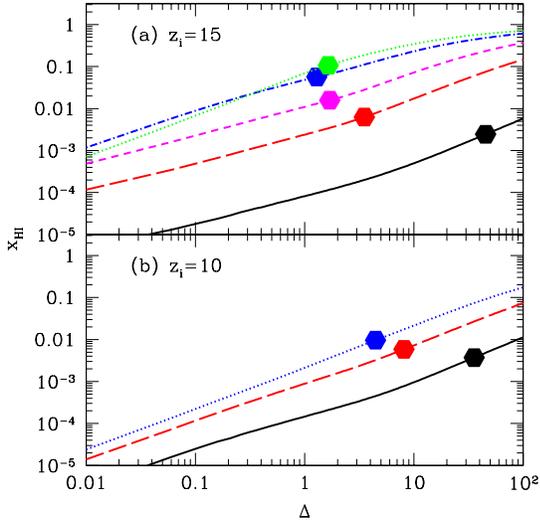}
\caption{Neutral fractions as a function of density at a sequence of redshifts, assuming that the ionizing background and mean free paths are set by the middle of the set of three curves in Fig.~\ref{fig:xicomp} .  \emph{Top:}  \emph{(a)}  The solid, long dashed, short dashed, dash-dotted, and dotted curves take $z=8,\,9,\,10,\,12$, and $14$, respectively.  \emph{(b)}  The solid, dashed, and dotted curves take $z=7,\,8$, and 9, respectively.  Note that all the $x_{\rm HI}$ curves ignore self-shielding.  The filled hexagons indicate where the MHR00 model predicts that self-shielding becomes important. }
\label{fig:xi_D_ib}
\end{figure}

The mean free path typically falls to $\lambda_i \sim 5$--$10 \Mpc$ before increasing again.  If we consider the uniform radiation field models shown in Figure~\ref{fig:mfp}, this is a rather large value and [for the $\bxion^g(z_i)$ assumed in this panel] the resulting ionizing background should easily maintain a high level of ionization in the gas.  This is important, because the gas \emph{between} the self-shielded regions must remain optically thin for the MHR00 model to apply.  To examine whether this is the case, we show in Figure~\ref{fig:xi_D_ib}\emph{b} the neutral fraction as a function of density at several stages during this recombination process, \emph{assuming a uniform ionizing background and neglecting self-shielding}.  The solid, dashed, and dotted curves show $\xhi(\Delta)$ at $z=7,\,8$, and 9, respectively (note that the entire universe is reionized by $z=6$).  We take $\bxion^g(z_i)=0.2$ for all the curves, so they correspond to the middle case in Figure~\ref{fig:xicomp}.  We have also used the mean free paths and emissivity from this model as input in the calculation, so it provides the actual $\xhi$ for all the gas that lies below the self-shielding threshold.  The filled hexagons show this threshold $\Delta_i$ according to the MHR00 model (and again taking the recombination history in Fig.~\ref{fig:xicomp}).

Obviously, even though the mean free path has declined significantly, the low-density gas remains quite highly ionized throughout the history.  At $z=9,\,8,$ and $7$, the volume-averaged neutral fractions are $\approx (1.9,\,  0.7,\, 0.14) \times 10^{-3}$ (including only gas with $\Delta < \Delta_i$), respectively.  Using equation~(\ref{eq:tau_simple}), these correspond to optical depths across one mean free path of $\tau_\lambda \approx (0.9,\,0.5,\,0.4)$, from high to low redshift.  

Because these optical depths are somewhat below unity, the MHR00 model appears to be a reasonable description of the IGM inside the fossil:  the accumulated opacity of gas \emph{between} self-shielded regions is small.  However, the absorption is not completely negligible, so there may be some pockets where the emissivity is smaller than normal (i.e., voids that are far from nearby galaxies), particularly in the earliest phases or when $\bxion^g(z_i) \la 0.1$, where the optical depths are somewhat larger.  To gauge whether this will occur, we must consider the scales over which the emissivity itself fluctuates significantly.

Before proceeding to do so, we note that Figure~\ref{fig:xi_D_ib} also lets us check another necessary condition for the MHR00 model:  the self-shielded gas must have had sufficient time to recombine and become neutral.  This is a particular worry for these fossils, because the typical densities of the self-shielded ``Lyman-limit systems'' are not too far above the mean (see the filled hexagons in Fig.~\ref{fig:xi_D_ib}).  Fortunately, the gas is indeed still able to recombine quickly:  the recombination time of (fully-ionized) gas with $\Delta=\Delta_i$ is, at worst, roughly on-tenth of the Hubble time.  Thus the self-shielded systems should be able to reach $x_i \sim 0.1$ without any trouble.  

Now we return to the question of whether isolated, low density regions will be able to break off and recombine more quickly.  A necessary, but not sufficient, condition to avoid returning to a swiss--cheese morphology and to maintain a uniform ionizing background is $\lambda_i > \bar{d}_{\rm gal} \equiv \bar{n}_{\rm gal}^{-1/3}$, where $n_{\rm gal}$ is the mean number density of galaxies.  If we include all halos with $T_{\rm vir} > 10^4 \kel$, $\bar{d}_{\rm gal} \approx 0.6,\,0.7,$ and $0.85 \Mpc$ and $z=6,\,8,$ and $10$, respectively.  In this example, $\lambda_i$ remains well above this level, and ionizing photons will, on average, be able to reach neighboring galaxies.  Thus we would not expect the most extreme swiss-cheese topology.  

However, inhomogeneity can actually begin before $\lambda_i$ reaches the mean galaxy spacing, because galaxies are highly clustered at high redshifts.  A convenient way to include clustering is to compare $\lambda_i$ to the characteristic size $R_c$ of ionized bubbles around clumps of galaxies in the background universe.  We compute $R_c$ as a function of ionized fraction using the analytic model of \citet{furl04-bub}, which is accurate to within a factor of $\sim 2$ during most of reionization, and better in the later stages \citep{zahn07-comp, mesinger07}.  We show $R_c$ by the dashed curves in the bottom panels of Figure~\ref{fig:xicomp} (see also \citealt{furl05-charsize}).\footnote{We cap $R_c$ at the mean free path when $\Delta_i=50$ for consistency with the fossil treatment. This roughly simulates the appearance of self-shielded regions in the background universe; see \citet{furl05-rec} for a more complete model.}

We see that $\lambda_i > R_c$ throughout the initial decline and plateau in $\bxion$.  This means that source clustering will be relatively unimportant while the bubble is recombining; the typical scale of variation is smaller than the residual mean free path in the bubble, so fluctuations in the ionizing sources will be washed out by the long mean free path (as in the post-reionization universe).  On the other hand, stochastic fluctuations in the galaxy distribution could still cause some inhomogeneity through a ``runaway'' effect, in which a region that happens to see a lower flux will become more neutral, which will help to shield neighboring gas parcels, leading to lower fluxes, etc.  Such a process would require $\tau_\lambda >1$, which is certainly possible in the earliest stages, or if $\bxion^g(z_i) \la 0.1$.  So it seems likely that some rare regions far from galaxies will begin to recombine.  In practice, this runaway effect will probably be reduced by the fact that under-illuminated regions are already the farthest from the galaxies and so are most likely to remain outside the ionized bubbles anyway.

Once $\bxion$ begins to increase, the left panel in Figure~\ref{fig:xicomp} shows that $\lambda_i \approx R_c$.  (This near equality is coincidental; see the right panel for a counter-example.)  At this stage, source clustering within the bubble \emph{will} become important.  In fact, even though the characteristic bubble size in the background universe is comparable to or only slightly larger than $\lambda_i$, the ``sub--bubbles'' within the fossil will actually be somewhat larger than in the surrounding universe. 

To understand this, note that $\bxion$ reaches a plateau when the emissivity is barely enough to overcome the modest recombination rate inside the bubbles (we find $C \approx 1$ at these times).  The fossil is then in a ``photon-starved'' regime where a larger emissivity is required in order to ionize additional hydrogen atoms.  Once $R_c \approx \lambda_i$, overdense clumps of sources are able to do just that over the scale of the mean free path in the background universe.  But in the fossil, only $\la 20\%$ of the hydrogen atoms are actually neutral -- so a clump of sources in the fossil can ionize a volume $\sim 5$ times larger than an equivalent clump in the background universe.  The actual sub--clump size may even be somewhat larger than this naive prediction, because the new, larger sub--bubble can merge with its neighbors as well.  On the other hand, the quasar is also likely to be inside an overdense region, which would decrease the expected size, at least in the central, biased parts of the fossil \citep{lidz07}.  We also note that the actual size of the galaxy sub--bubbles can be rather large (several Mpc at $z\lsim 8$), so that a typical fossil may contain only a few dozen such sub--bubbles.

Meanwhile, if these highly-ionized regions around clumps of sources are able to grow because their emissivity is larger than average, we also know that underdense voids within the fossil will continue recombining even during the phases where $\bxion$ is either flat or increasing.  Thus a contrast between the fully ionized regions and their recombining neighbors will develop, and the fossil will look much like the background universe but with larger bubbles (by a factor $\sim 1/\bxion$ in volume) and a significantly smaller contrast between the fully ionized regions and the surrounding partially neutral gas.  Even these shielded regions would not provide much contrast because much of the dense gas has already recombined, so the effective clumping is modest.  We emphasize that, even in the absence of \emph{any} ionizing sources, fossils recombine quite slowly in their late stages (see Fig.~\ref{fig:fossil}), and only a modest emissivity is required to halt the decrease.  (For example, the three solid curves span a factor of four in emissivity -- much larger than we expect from large-scale density fluctuations in the IGM -- and still the plateau in $\bxion$ occurs at a quite large value.)

\subsection{An Early Quasar} \label{early}

Quasars at $z_i=15$ will produce fossils with more dramatic contrasts, because reionization at $z \sim 6$--$10$ implies $\bxion^g(z_i) \la 0.02$.  Thus there is a long time for the bubble to remain visible against the mostly neutral background (and, eventually, to recombine).  However, even in this extreme case, the right panels of Figure~\ref{fig:xicomp} show that the bubbles never become more than about one-half neutral before the background catches up to them.  One interesting side note is that in this high--redshift case, the MHR00 clumping factor implies \emph{more} residual neutral gas; as seen in Figure~\ref{fig:clump}, $C \la 1$ except in the initial stages, because only underdense gas (where the recombination time is long) remains ionized after the initial phases.  (At $z \ga 14$, the MHR00 clumping factor does lead to slightly faster recombinations, although that is difficult to see in the figure.)

The degree of recombination in these bubbles also suggests that the MHR00 prescription may break down.  Figure~\ref{fig:xi_D_ib} confirms this explicitly for the $\bxion^g(z_i)=0.01$ case (the middle curves in the right panels of Figure~\ref{fig:xicomp}).  Here we find that $\Delta_i \approx 1$ for most of the evolution, which leads to mean free paths $\sim 1 \Mpc$.  (At this extremely high redshift, the IGM is still fairly uniform, so there is little variation of the mean free path with the ionized fraction; see Fig.~\ref{fig:mhrmfp}.  Thus $\lambda_i$ falls quickly to $\sim 1 \Mpc$ and then levels off for a long period, including the plateau and initial increase in $\bxion$.)  Until the end stages of reionization, the ionized fraction in the $\Delta < \Delta_i$ gas (nominally optically thin in the MHR00 picture) is a few percent at $z \ga 10$, so $\tau_\lambda \approx 13,\,4.5,\,1.5,\,0.7,$ and $0.1$ at $z=14,\,12,\,10,\,9$, and $8$, respectively.  Thus, \emph{even assuming uniform illumination}, the IGM recombines sufficiently to render it opaque to ionizing photons on Mpc scales.  The MHR00 picture breaks down, and the mean free path will become a fraction of an Mpc.

Of course, luminous sources will be able to maintain high ionization in their own environs, while more distant gas will continue to recombine.  The mean spacing between galaxies at $z=15$ (10) is $2.1$ $(0.85) \Mpc$, respectively.  Comparing to Figure~\ref{fig:xicomp}, $\lambda_i$ falls below this level, at least until $\bxion^g \la 0.5$, even using the overestimate from the MHR00 picture.  Thus the early stages of the recombination process \emph{must} be highly inhomogeneous:  regions surrounding rare galaxies will remain highly ionized, but other regions will recombine much like the cases shown in Figure~\ref{fig:fossil}.  As a result, the ionized fraction inside of these fossils will be \emph{smaller} than shown in Figure~\ref{fig:xicomp}, because much of the gas will be recombining without interference while the ionizing sources waste photons repeatedly ionizing dense gas in their environs.  This could lead to much higher contrast between the ionized bubbles and partially ionized surroundings than is ever possible for a fossil surrounding a $z_i=10$ quasar.

Once the clustering of the ionizing sources becomes significant, $R_c$ quickly becomes larger than $\lambda_i$ (again, even with the overestimate from the MHR00 picture).  This means that source clustering will induce stronger variations in the fossil's ionization structure that begin at a smaller $\bxion^g$ (compared to the $z_i=10$ fossil).  Again, the ionization pattern inside the fossil will have a larger characteristic scale than that in the background universe, because the surrounding gas has less neutral hydrogen, but the contrast between the fully ionized bubbles and the rest of the IGM will be somewhat smaller than in the background universe.  

Overall, we find that only in extreme cases like this one -- where the quasar shuts off while the background universe is only a few percent ionized -- do we expect a well-developed swiss-cheese topology inside the fossil.  Attaining such a state is difficult for two reasons.  First, the low-density gas requires significant time to recombine; because $\ga 50\%$ of the mass has $\Delta \la 1$, the quasar must appear early: at least several recombination times before the completion of reionization. Furthermore, a swiss-cheese topology also requires that the mean free path within the fossil falls below $R_c$ (or, even better, the mean galaxy spacing) so that the clustering pattern of the galaxies can induce fluctuations in the emissivity and hence ionized bubble pattern.  From Fig. \ref{fig:xicomp}, we see that $\lambda_{i}$ grows more steeply with $\bar{x_{i}}$ than does $R_{c}$; only if significant recombination takes place does $\lambda_{i}$ fall below $R_{c}$. 

\begin{figure*}
\plottwo{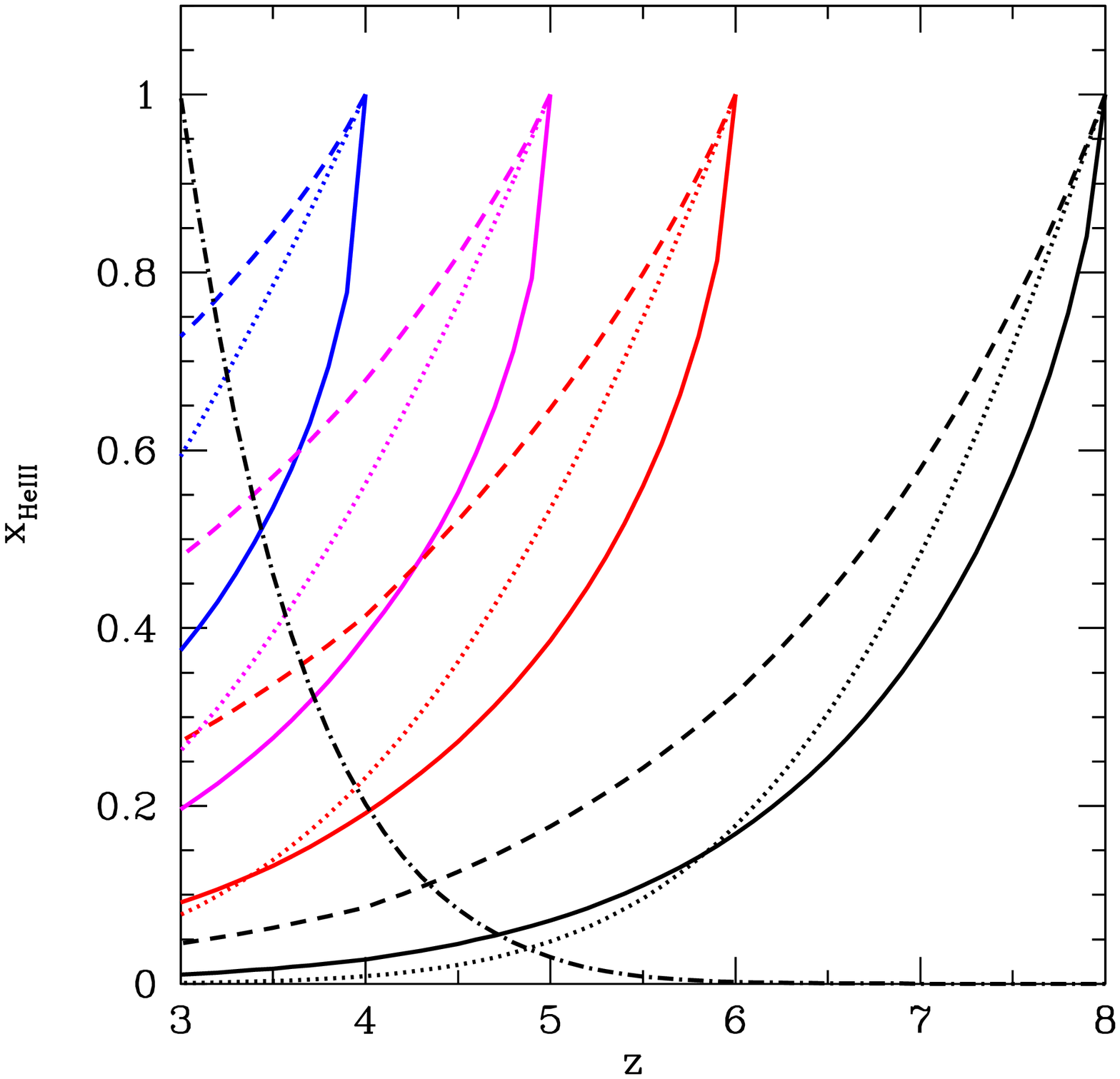}{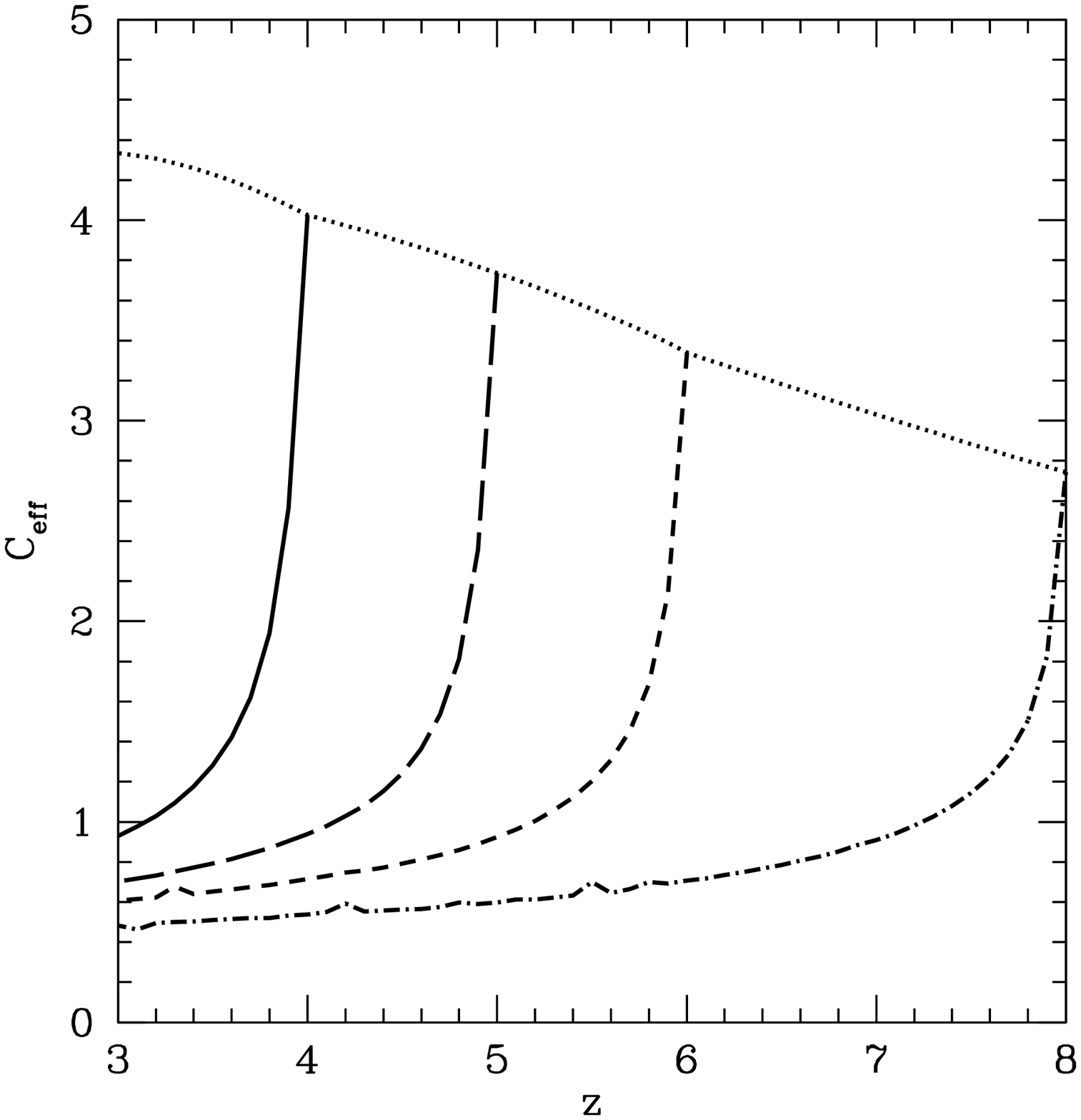}
\caption{\emph{Left panel:} Ionization histories for fossil HeIII regions with zero ionizing background. 
The four sets of curves that decrease from right to left assume $\bar{x}_{\rm HeIII}=1$ at $z_i=4,\,5,\, 6,$ and $8$.  Within each set, the solid and dashed curves show $\bar{x}_{{\rm HeIII},m}$ and $\bar{x}_{{\rm HeIII},v}$, respectively.  The dotted curves show the histories for elements with $\Delta=1$.  The dot-dashed curve (increasing from the right) shows a sample global reionization history.  \emph{Right panel:}  Same as Figure~\ref{fig:clump}, but for helium reionization.}
\label{fig:helium}
\end{figure*}

\subsection{Residual Emission from the Quasar} \label{post-quasar}

Thus far, we have assumed a ``lightbulb" model of quasar evolution, in which quasars only turn on for some short period of time, and then shut off, leaving only the remaining galaxies to supply an ionizing background to suppress recombinations. However, recent models motivated by hydrodynamical simulations of galaxy mergers -- in which black holes spend the majority of the time well below the peak luminosity of the associated quasar, but nevertheless radiating -- provide better fits to multi-wavelength quasar luminosity functions and a host of other empirical constraints \citep{hopkins05b, hopkins05c}. If the quasar continues shining at some lower luminosity, it may suffice to prevent the bubble from recombining, even without the aid of neighboring galaxies. If the quasar shines at $L_{\rm hi}$ for time $t_{\rm hi}$ and then reverts to $L_{\rm lo}$ for an extended period, the requirement that the bubble does not recombine during the latter period is $ (L_{\rm lo}/L_{\rm hi}) \gsim (t_{\rm hi}/t_{\rm rec}) \sim 0.03 \, (t_{\rm hi}/10^{7} {\rm yr}) \, C \, ([1+z]/10)^{3}$, where $t_{\rm rec} = 1/[C \alpha_{\rm A} n_{e}]$ is the recombination time within the bubble.  This accords well with expectations from the \citep{hopkins05b} model, in which quasars shine on cosmological timescales with Eddington ratios between $l \sim 0.01$ and $l \sim 1$ at peak. A fit to their simulations which also agrees with fits to luminosity functions is:
\begin{equation}
{\deriv t \over \deriv {\rm log} \, L}= t_{\rm Q}^{*} {\rm exp} \left( -L/L_{\rm Q}^{*} \right),
\end{equation}
where the fitted parameters $L_{\rm Q}^{*}=\alpha_{\rm L} L_{\rm peak}$, $t_{\rm Q}^{*}= t_{*}^{(10)} (L_{\rm peak}/[10^{10} \, L_{\odot}])^{\alpha_{T}}$, 
and $\alpha_{\rm L}=0.20$, $t_{*}^{(10)}= 1.37 \times 10^{9}$~yr and $\alpha_{T}=-0.11$. This light curve allows us to calculate the radiative output as a function of luminosity.\footnote{Strictly speaking, in the simulations motivating the \cite{hopkins05b} fitting formula, a somewhat larger fraction of the low-Eddington ratio emission is produced {\it before}, rather than after the bright, near-Eddington phase.}  For instance, although most of the energy is radiated (and hence most of the bubble growth occurs) while the quasar is at peak luminosity, a $L_{\rm peak} \sim 10^{10} \Lsun$ quasar spends $\sim 1$~Gyr with $L > 0.03 L_{\rm peak}$, sufficient to prevent the bubble from recombining until the universe is reionized. In this model, lower mass black holes radiate an even large fraction of energy at low Eddington ratios. Of course, the applicability of the \citet{hopkins05b} model to high redshifts is uncertain, but it is clear that low-luminosity quasar remnants could contribute very significantly to the ionizing background in 'fossil' bubbles. Hence, our arguments considering only the galactic contribution provide a strict lower bound on the ionization state.  

\section{Helium Reionization} \label{helium}

\subsection{Zero Ionizing Background} \label{he-zero}

Figure~\ref{fig:helium} shows corresponding recombination histories for fossil HeIII bubbles produced during quasar reionization.  We consider four initialization redshifts here, $z_i=4,\,5,\, 6,$ and $8$, from left to right, and assume zero ionizing background.  Qualitatively, the results are similar to those for HI:  the mass-averaged ionized fraction is typically smaller than the ionized fraction for gas at the mean density, but not by a large amount.  However, once most of the gas has recombined, $\bar{x}_{{\rm HeIII},m}$ becomes slightly \emph{larger} than the ionized fraction of gas with $\Delta=1$:  beyond this point, most of the neutral gas lies in underdense voids, and the effective clumping factor is substantially \emph{less} than unity.  The curves are also somewhat steeper than for hydrogen because the recombination time for helium is several times smaller and because the Universe is clumpier at lower redshifts.

Figure~\ref{fig:helium} also shows that the clumping factor behaves similarly to fossils during hydrogen reionization (although here there is no initial spike, because the temperatures are assumed to be hotter at the initial instant of reionization).  The asymptotic $C_{\rm eff}$ are actually smaller here, mostly because the recombination rate is slightly faster relative to the Hubble time and so only gas below the mean density remains substantially ionized.

As before, it is useful to compare $x_{\rm HeIII}$ within bubbles to the background universe.  The dot-dashed curve in the left panel of Figure~\ref{fig:helium} shows such a history (see \citealt{furl07-helium} for details).  It uses the quasar luminosity function from \citet{hopkins07}, assumes that quasars have spectra typical of observed quasars at lower redshifts \citep{vandenberk01, telfer02} and ignores IGM recombinations (so is no more than a rough guide).  We have reduced the total luminosity of each quasar by a factor of $\sim 2.5$ to force reionization to complete at $z=3$, consistent with a number of observational probes (see \citealt{furl07-helium} for a summary).  Interestingly, the background ionized fraction increases extremely rapidly, with most of the ionizations occurring at $z \la 4$. 
On the one hand, this means that most bubbles form not long before reionization is complete and have relatively little time to recombine.  On the other hand, the fossils that do form around the rare luminous quasars at $z \ga 5$ have an extremely high contrast with the background universe and have ample time to recombine nearly fully.

\subsection{A Non-zero Ionizing Background} \label{nzion-helium}

We have so far assumed that fossil helium bubbles evolve in isolation, without being exposed to any HeII--ionizing radiation.  In contrast to hydrogen reionization, that is not a bad assumption -- at least for a time -- because quasars are so rare, and galaxies (probably) do not contribute significantly to helium reionization.  Thus, once a bubble appears, it will remain free of any ionizing sources until another quasar forms in the same region.  The key question is how long this empty phase will last.

To address this question, we must first determine what sphere of influence a new quasar will have in the partially ionized medium.  $L_\star$ quasars (with lifetimes of $10^7 \yr$) can ionize all the helium inside regions within $\sim 15 \Mpc$ of their host galaxy.  Once regions are mostly ionized, so that battling recombinations presents the main challenge (rather than ionizing new material), \citet{furl07-helium} showed using the MHR00 model that their light is attenuated only over $\sim 35 \Mpc$ distances.  Thus the relevant question is the time lag before a quasar appears in this maximum volume $V_{\rm max} \sim 30 \Mpc$.  To estimate this, we write the quasar number density as $n_{\rm QSO} = 10^{-6} n_{-6} \Mpcden$ and the lifetime as $t_{\rm QSO} = 10^7 t_7$ yrs.  For the purposes of a simple estimate, we then assume that the quasar population over the time lag $\Delta t$ can be divided into $N$ discrete generations.  The number of generations required to have a probability of one-half to find a second source in the region is then $\sim (2 n_{\rm QSO} V_{\rm max})^{-1}$, or
\bq
H(z) \Delta t \sim 0.047 {t_7 \over n_{-6}} \, \left( {1+z \over 5} \right)^{3/2}.
\label{eq:hegen}
\eq
According to the \citet{hopkins07} model, the number density of quasars with $L \ga 10^{11.5} \Lsun$ (roughly $L_\star$) is $n_{-6} \sim (2.2,\,0.73,\,0.16)$ at $z=(3,\,4,\,5)$, implying $H(z) \Delta t \sim (0.02,\,0.06,\,0.4) t_7$ at these redshifts.  Thus once reionization gets underway, the time lag between successive generations of ionizing sources is rather small -- only if a region is ionized early in reionization will it have sufficient time to recombine significantly.  For example, a $z=4$ (5) bubble will reach $\bar{x}_{{\rm HeIII},m} \sim 0.6$ ($0.4$) before it is reionized by another quasar.  We therefore expect roughly one-half of all photons produced at $z \ga 4$ to be ``wasted'' through fossils.  But by $z=3.5$, the fossil lifetimes will be so short that few photons are wasted.

\section{Discussion} \label{disc}

\begin{figure}
\plotone{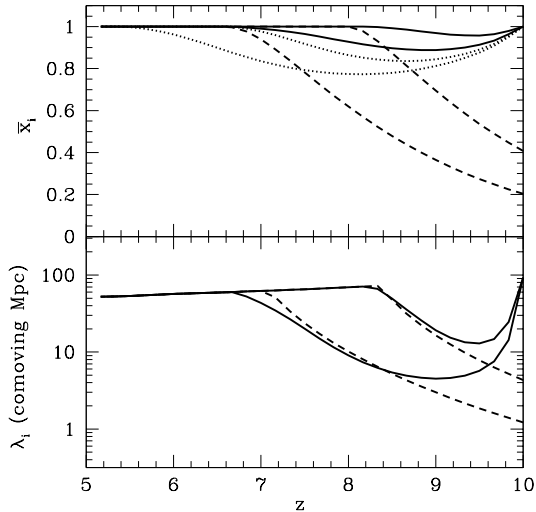}
\caption{Ionization histories inside fossils and in the background universe, as in Fig.~\ref{fig:xicomp} with $z_i=10$.  The solid and dashed curves are identical to the $\bxion^g(z_i)=0.2$ and 0.4 curves in Fig.~\ref{fig:xicomp} with $C=C_{\rm MHR}$.  The dotted curves also take $\bxion^g(z_i)=0.2$ and 0.4 with $C=C_{\rm MHR}$, but they assume that photoheating raises the Jeans mass by a factor of ten within the fossils.}
\label{fig:xicomp-fb}
\end{figure}

\subsection{Photoionization Feedback} \label{fb} 

As noted above, it is possible that photoheating increases the Jeans
mass in the fossil bubbles \citep{rees86, efstathiou92, thoul96},
and with it the effective minimum galaxy mass $\mmin$ over that in the
universe at large.  The efficiency of this feedback may be reduced
at redshifts much larger than $z\gsim 6$ \citep{dijkstra04-feed};
nevertheless, it can decrease the emissivity in fossils.  It is therefore
natural to ask whether such a reduction may allow fossils to recombine
more rapidly than we found above.

Figure~\ref{fig:xicomp-fb} shows a set of ionization histories that
address this issue.  The solid curves are identical to the
$\bxion^g(z_i)=0.2$ and $0.4$ curves in Figure~\ref{fig:xicomp}.  In
the dotted curves, we assume that photoheating inside the fossil
bubble has increased the minimum mass to $\mmin=10m_4$ \citep{rees86,
  efstathiou92, thoul96}.  Although photoheating decreases the effective collapsed fraction within the bubble by nearly an order of magnitude, the effect on $\bxion$ is relatively modest, and the region still remains highly ionized.  In part, this is because these massive galaxies evolve more rapidly, so the \emph{emissivity} (proportional to $\deriv \fcoll/ \deriv z$) does not fall by as large a factor as $\fcoll$ itself.  Note also that we still compare to a background universe with $\mmin=m_4$; this is why the fossil can have a smaller ionized fraction than the universe at large.  Of course, in reality photoheating will also slow down the background evolution once $\bxion^g \ga 0.5$, and the two will map more smoothly onto each other \citep{haiman03, furl05-double}.  But this simple treatment suffices to show that photoheating feedback will not allow fossils to recombine significantly more than in our fiducial cases; instead, they are roughly equivalent to a normal history with one-half the emissivity.

However, because massive galaxies are much rarer, inhomogeneity will be easier to re-establish.  In particular, $\lambda_i$ can be comparable to or smaller than the mean galaxy spacing if photoheating feedback suppresses galaxy formation inside of fossil HII regions.  If we only include galaxies with $T_{\rm vir} > 2 \times 10^5 \kel$ (i.e., maximal photoheating feedback), $\bar{d}_{\rm gal}=4.4,\,7.3,\,13.7,$ and $110 \Mpc$ at $z=6,\,8,\,10,$ and $15$.  Comparing to Figure~\ref{fig:xicomp-fb}, the mean free path inside these bubbles can fall below $\bar{d}_{\rm gal}$ in moderately ionized volumes, even at relatively low redshifts.  However, the effect of photoheating feedback is probably smaller during or just after a region is reionized, because objects already in the midst of collapse are not very susceptible to it \citep{dijkstra04-feed}.  Thus it is unlikely that suppression at this level can be achieved in the relatively short time interval between $z=10$ and the end of reionization.  We would nevertheless generically expect more inhomogeneity inside the bubbles if photoheating feedback occurs.

\subsection{Detecting Fossil Bubbles} \label{detect}

We have now established that most fossil bubbles formed in the midst of hydrogen reionization will remain highly ionized throughout the entire process.  We will now consider strategies to find them.  To do so, we must rely on two features of fossils:  their large size, and possibly a different topology of ionized gas.  We have already seen that the latter effect is not so important; most bubbles will remain highly ionized, and those that do begin to recombine significantly will resemble the background universe but with larger bubbles and less contrast -- which will probably only be distinguished with 21 cm survey instruments well beyond those currently being built, such as the Square Kilometer Array.  Instead, it is their large sizes that are often taken to be the distinguishing characteristic of fossils.

The first potential problem is distinguishing quasar-driven fossils from large ionized regions generated by ``normal'' star formation during reionization.  As illustrated in Figure~\ref{fig:xicomp}, \citet{furl04-bub} established that \htwo regions inevitably reach extremely large sizes during reionization (see also \citealt{furl05-charsize}).  During most of the process, the characteristic size is relatively small in comparison to the ionized regions generated by luminous quasars (which are $\sim 20$--$40 \Mpc$ across, assuming typical quasar parameters).  But this is only the characteristic size -- in reality, the bubble distribution has a tail of rare, large objects.  Because the quasars themselves are also extremely rare, it is not obvious which type of rare large bubble will outnumber the other.

\begin{figure}
\plotone{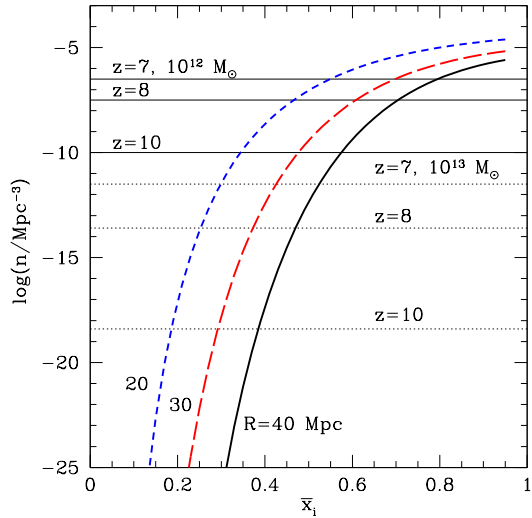}
\caption{Comparison of the number densities of massive halos (assumed
  to have hosted luminous quasars) and large bubbles from ``normal''
  reionization (dominated by stellar sources).  The thick solid,
  long-dashed, and short-dashed curves show the abundance of regions
  with $R>40,\,30,$ and $20 \Mpc$, respectively, from stellar
  reionization, based on the model of \citet{furl04-bub}.  These are
  evaluated at $z=8$ but only depend on redshift slightly.  The
  horizontal dotted and solid lines show the total abundance of halos
  with $m>10^{13} \Msun$ and $10^{12} \Msun$, respectively, at the
  specified redshifts, representing possible quasar hosts.}
\label{fig:nbub-halo}
\end{figure}

Figure~\ref{fig:nbub-halo} illustrates the problem quantitatively.  The three thick curves give estimates for the number density of large ionized bubbles from normal star formation.\footnote{In detail, we compute $x_i(>R)/(4 \pi R^3/3)$, where $x_i(>R)$ is the fraction of space filled by bubbles larger than $R$.  This is more robust to uncertainties in the maximum bubble size set by recombinations, which breaks extremely large \htwo regions into smaller sections \citep{furl05-rec}.}  We show $R=20,\,30,$ and $40 \Mpc$, from top to bottom.  We calculate these with the analytic model of \citet{furl04-bub}; numerical simulations show that this model underestimates the characteristic bubble size early in reionization (by a factor of a few), but is quite accurate in the later stages most applicable to this argument \citep{mesinger07}.  

The thin horizontal lines show the number density of halos above mass thresholds $10^{12} \Msun$ and $10^{13} \Msun$ (solid and dotted lines, respectively) calculated from the \citet{sheth99} mass function at $z=7,\,8,$ and 10 (from top to bottom within each set).  If we assume that these massive halos all host supermassive black holes, then the number density of halos above that threshold will equal the maximum number density of fossils as well (ignoring recombinations).  The luminous $z=6$ quasars are thought to lie in halos of $m \sim 10^{12}$--$10^{13} \Msun$ from both number density arguments \citep{fan00,haiman01-qso} and spectral signatures in the optical \citep{barkana03-infall} and molecular line spectra \citep{walter04, narayanan08}.  This mass range is also consistent with converting the empirically-estimated black hole masses \citep{vestergaard04} to a halo mass using the black hole mass-halo mass and black hole mass-velocity dispersion relations calibrated at lower redshifts \citep{magorrian98, ferrarese00, gebhardt00}.  

Comparing the bubble and halo number densities, it is clear that fossils will not necessarily be distinguishable from more normal (galaxy--driven) ionized zones past about the midway point of reionization.  Evidently, the rare tail of ionized bubbles may provide a comparable number of objects to the quasars themselves.  That said, the uncertainties in both the tail of the bubble distribution (which depends on the overlap of smaller regions) and the conversion from quasar luminosity to halo mass (and hence number density) are highly uncertain, and it is possible that fossils will, in fact, stand out relatively well.  There are also two morphological effects that could be helpful (although difficult to interpret robustly).  First, the large halos hosting quasars are also surrounded by substantial clusters of sources, so they already sit in large ionized bubbles from which the quasar can build a somewhat larger ionized region \citep{lidz07}.  Second, quasars may present distinctive ionization patterns (perhaps two collinear cones, if the emission is beamed, or a spherical region if it is isotropic).

Furthermore, high-redshift galaxy surveys over the same fields may allow us to break this degeneracy. 21 cm and galaxy surveys should have a strong detectable anti-correlation \citep{furl06-cross,wyithe06-cross}. Moreover, on the largest scales where bubbles are directly detectable, we could simply compare the observed comoving emissivity within the bubble to its size.\footnote{Note that the bubble size is proportional to the integrated star formation history $\propto f_{\rm coll}$, rather than the instantaneous comoving emissivity $\propto \dot{f}_{\rm coll}$. However, the two are strongly correlated in the sense that the rarest high density regions with the highest $f_{\rm coll}$ will also have the highest rate of collapse.} This relation will have some natural scatter due to the past star formation history within the bubble, variations in gas clumping, and observational error. However, since quasars constitute a strong impulsive spike in the comoving emissivity, fossil quasar bubbles will induce a decorrelation between the observed comoving emissivity and bubble size. In particular, quasar fossils will lie far off the tail of the emissivity distribution for a given bubble size (with the deviation being proportional to the fraction of ionized volume contributed by the quasar).  Such effects could potentially allow us to distinguish quasar-blown bubbles (the majority of which will be fossil regions), permitting a lower bound on the contribution of quasars to reionization. 

Another potential problem in identifying these bubbles is that the underlying ionization (and density) fields are fluctuating on large scales from the cumulative effects of smaller regions.  We need to be able to distinguish the fossil region's signal from this background ``confusion noise''.  The best way to discover fossil bubbles will probably be through the 21 cm line \citep{furl06-review}, for which the brightness is proportional to the total density of neutral gas.  A fully ionized fossil would have a contrast $\sim 20 \bxhi \mkel$ with the mean background.  To estimate the confusion limit in such an observation, we use the simulations of \citet{mesinger07} to measure the r.m.s. fluctuation amplitude at wavenumbers $k \sim (2 \pi/30) \Mpcinv$.  This varies from $\sim 0.1$--$1 \mkel$ over the range $\bxhi \sim 0.8$--$0.1$.  Thus we would expect fossils to stand out very well -- many standard deviations from the mean -- except near the end of reionization, when they will begin to blend into the background fluctuations as well.  The main obstacle toward direct detection is therefore telescope noise, though sufficiently large bubbles may still be detectable, even by the first generation of instruments, at least at low redshifts (see \S 8.6 of \citealt{furl06-review}).  

Although the uncertainties in these estimates are fairly large (especially in the abundance of massive quasars), these factors do suggest that it may be significantly more difficult to cleanly identify fossil bubbles than often assumed, at least late in reionization:  they may not be much more abundant than large bubbles from star formation, they may not stand out strongly from the background field, and their ionization pattern (either remaining highly ionized or with discrete ionized regions separated by partially ionized walls) does not qualitatively differ from the background.  We suspect that some modeling will be required to distinguish the two cases.

\subsection{Fossil and Active Bubbles} \label{active}

As we have seen above, although the fossils can be kept highly
ionized, they are not fully ionized, and will thus appear different
from active quasar bubbles (for example, in 21 cm imaging, they will
have a somewhat reduced contrast).  Since active quasars, with the
luminosities required to produce large ($\gsim 10$ Mpc) fossil
bubbles, should be relatively easily detectable even at $z>6$, active
bubbles can probably be distinguished routinely from fossil bubbles,
simply by the presence or absence of a luminous ionizing source.

We now turn to the question of how many fossil bubbles we might expect relative to those surrounding active quasars.  Fossils forming relatively close to the end of reionization (at $z \la 10$ for reionization at $z \sim 7$) do not significantly recombine so would appear as mostly empty fossils long after their source quasar fades.  This will boost the number of large (mostly) ionized regions well above that expected from the quasar luminosity function alone, and can potentially allow one to constrain the quasar lifetime \citep{wyithe04-qso, wyithe05-qso}.

However, because fossils persist for so long, the quasar abundance itself probably evolves significantly over the recombination time.  As an example, let us suppose that the number density of bright quasars follows $n_{\rm QSO} \propto (1+z)^{-\beta}$, with $\beta>0$, and that fossils remain ionized so long as they form at $z<z_{\rm max}$.  In that case, assuming that the quasar lifetime $t_{\rm QSO}$ is much smaller than the elapsed time from $z_{\rm max}$ to the observed time $z_{\rm obs}$, we expect
\bq
n_{\rm fossil}(z_{\rm obs}) \sim {n_{\rm QSO}(z_{\rm obs}) \over (\beta+3/2) H(z_{\rm obs}) t_{\rm QSO}}.
\label{eq:nfossil}
\eq
The number density of bright quasars has $\beta \sim 4$ over the range $z \sim 3$--$6$ \citep{fan01-lf,fan04}.  If this continues to higher redshifts, it will provide a factor $\sim 5$ suppression from the naive value.  Unless it can be measured independently, this evolutionary factor will pose a significant (factor of several) uncertainty in the measurement of $t_{\rm QSO}$ from the fossil to active ratio.  One way to break the degeneracy is by measuring the luminosity function over the entire time interval -- but doing so would require a search for active quasars that extends to much higher redshifts than the fossil search itself.
Another potential method would be a series of measurements of the number density of bubbles over a redshift interval comparable to the recombination time of a bubble. This could probe the redshift evolution of the quasar luminosity function, particularly if sensible priors for the quasar lifetime are adopted. 
Finally, we note that the shapes of active quasar bubbles could, in
principle, contain information on the host quasars and the IGM,
through finite light--travel time effects \citep{wyithe04-qso,yu05a}.
The results we obtained here, i.e. that most of the fossil gas takes
significantly longer to recombine than one would estimate using a
fiducial IGM clumping factor, will hamper statistical versions of this
measurement \citep{sethi08}, since fossils will outnumber active
quasar bubbles and not show the apparent distortion due to finite
light travel time.

\subsection{Double Reionization} \label{double}

Up to this point, we have studied recombinations in the context of fossil bubbles, but much of our discussion also applies to so-called ``double reionization'' scenarios \citep{cen03, cen03-letter, wyithe03, wyithe03-letter} in which a first generation of massive, metal free stars reionizes most of the universe but also enriches it -- inducing a transformation to much less efficient, but more normal, stars.  In the scenarios originally proposed, photons from the second generation did not suffice to battle recombinations and so the universe became substantially (up to 80\%) neutral, until structure formation had progressed enough for the more normal stars to complete reionization.

However, these models rely on relatively large clumping factors to speed up recombinations so that a recombination era can occur during the relatively short interval between the two ionizing source generations.  Reionization requires that (at least) one ionizing photon per hydrogen atom be produced; this is relatively difficult at high redshifts.  But \emph{maintaining} a high ionization fraction is less stringent:  it requires only that one ionizing photon be produced per recombination.  For gas at the mean density, where the recombination time is roughly the Hubble time, this is relatively easy.  We have seen that the effective clumping factor declines as more gas recombines, suggesting that the ``recombination era'' will probably be confined to dense pockets of gas.

Our models strengthen this conclusion.  Figure~\ref{fig:xicomp} can be used to analyze double reionization scenarios in addition to fossil bubbles:  we can simply think of the $\bxion^g(z)$ curves as describing the second generation, and we suppose that the fossil's initialization redshift describes the moment at which the first generation shuts off.  The left and right panels then mimic moderate and more extreme double reionization scenarios.  Our most important conclusion is that, even in the  most extreme case -- when the second generation is able to achieve only $\bxion \sim 0.005$ at the initial point -- less than half of the gas is actually able to recombine.  In more moderate scenarios, only $\sim 10\%$ can recombine, and that is generally dense gas surrounding galaxies and halos.

Note finally that these scenarios assume a sudden transition between the two generations -- in reality, that is much too simplistic, because all of the feedback mechanisms that can mediate such a transition (metal enrichment, photoheating, etc.) act locally and on relatively long time scales \citep{haiman03, dijkstra04-feed, furl05-double, iliev07-selfreg, mcquinn07}.  Both of these effects (slow recombinations and slow transitions) will combine to make double reionization exceptionally unlikely.

\section{Summary} \label{summ}

We have examined recombinations in inhomogeneous fossil ionized
regions, adapting the simple MHR00 model of the density distribution
of the IGM, and tracking the ionization histories of individual gas
elements.  Without a residual ionizing background, the process occurs
quickly at first, because the dense IGM gas can recombine rapidly
(during both hydrogen and helium reionization).  But once these
relatively rare pockets become neutral, recombinations slow
dramatically, and eventually $C_{\rm eff} \la 1$.  Thus it is quite
difficult for a region to recombine entirely, even in the absence of 
any ionizing flux.

The ionizing background from galaxies inside the fossil,  together with any residual low--level emission from the black hole past its bright quasar phase, efficiently further suppresses recombinations during hydrogen reionization. So long as galaxies are able to ionize $\ga 20\%$ of the universe on average, they will be able to halt recombinations in all but the densest gas inside the fossil regions.  In most cases, the mean free path of ionizing photons remains large enough to wash out any fluctuations in the ionizing source population, and the fossil remains uniformly ionized (outside of self-shielded regions).  In more extreme scenarios, when the fossil is created long before reionization, the gas can recombine significantly (down to roughly the mean density), which will probably allow large regions far from galaxies to shield themselves and begin recombining faster.  In such a case, the fossil IGM will look qualitatively similar to the swiss--cheese ionization topology of the IGM elsewhere, but with larger ionized regions and a smaller contrast between the ionized and ``gray'' partially neutral regions.  

These fossil regions may therefore be somewhat harder to distinguish from the normal IGM than often assumed -- more so because ``normal'' ionized regions can reach comparable sizes to those of quasar zones near the middle of reionization, and may be as or even more abundant (albeit both of these estimates rely on the rare tails of theoretical distributions, in galaxies and ionized bubbles, and have very large uncertainties).  

During helium reionization, fossil bubbles evolve somewhat differently.  Because the ionizing sources are rare, once a region's source quasar shuts down, it is unlikely to be illuminated by any other source for a time.  This will allow the gas to recombine without interference; because the universe is so clumpy at $z \sim 4$, a large fraction of the gas can recombine over that interval.  However, the typical time lag before the next quasar appears is only a fraction of the Hubble time, so in practice less than half of the gas will be able to recombine, even for quasar bubbles formed early in the reionization process.

\acknowledgments

This research was partially supported by the NSF through grants AST-0607470 (SRF) and AST-0407084 (SPO) and by NASA through grants NNG04GI88G (ZH) and NNG06H95G (SPO). ZH
also acknowledges support by the Pol\'anyi Program of the Hungarian
National Office of Technology.


\end{document}